\pdfoutput=1
\documentclass[twocolumn]{autart} 
\usepackage{cite}
\usepackage{amsmath,amssymb,amsfonts,bbm,epsfig}
\usepackage{subfiles}
\usepackage{graphicx,xcolor}
\usepackage{textcomp, hyperref}
\usepackage{tikz, verbatim, array}
\usepackage[linesnumbered,ruled,vlined,norelsize]{algorithm2e}

\newcolumntype{P}[1]{>{\centering\arraybackslash}p{#1}}
\newcommand\rev[1]{{\color{black}{#1}}}
\newcommand\revv[1]{{\color{black}{#1}}}
\SetCommentSty{mycommfont}
\SetKwComment{Comment}{$\backslash \backslash$\ }{}

\begin{document}

\begin{frontmatter}
\title{Maintaining Ferment: On Opinion Control Over Social Networks\thanksref{footnoteinfo}} 
\thanks[footnoteinfo]{This work was done when Mohak Goyal was with IIT Bombay. A preliminary version of this work was presented at the IEEE CDC 2019~\cite{goyal2019maintaining}}
\author[SUMSE]{Mohak Goyal}\ead{mohakg@stanford.edu},
\author[IITBEE]{Nikhil Karamchandani}\ead{nikhilk@iitb.ac.in},    
\author[IITBSC]{Debasish Chatterjee}\ead{dchatter@iitb.ac.in},               
\author[IITBEE]{D. Manjunath}\ead{dmanju@iitb.ac.in}  

\address[SUMSE]{Department of Management Science \& Engineering, Stanford University, USA}  
\address[IITBEE]{Department of Electrical Engineering, Indian Institute of Technology Bombay, India}             
\address[IITBSC]{Systems \& Control Engineering, Indian Institute of Technology Bombay, India}        
          
\begin{abstract}                          
We consider the design of external inputs to achieve a control
  objective on the opinions, represented by scalars, in a social
  network. The opinion dynamics follow a variant of the discrete-time
  Friedkin-Johnsen model. We first consider two minimum cost optimal
  control problems over a finite interval $(T_0,T),$ $T_0 >0$---(1)~TF
  where opinions at all nodes should exceed a given $\tau,$ and (2)~GF
  where a scalar function of the opinion vector should exceed a given
  $\tau.$ For both problems we first provide a Pontryagin maximum
  principle (PMP) based control function when the controlled nodes
  are specified. We then show that both these problems exhibit the
  turnpike property where both the control function and the state
  vectors stay near their equilibrium for a large fraction of the
  time. This property is then used to choose the optimum set of
  controlled nodes. We then consider a third system, MF, which is a
  cost-constrained optimal control problem where we maximize the
  minimum value of a scalar function of the opinion vector over
  $(T_0,T).$ We provide a numerical algorithm to derive the control
  function for this problem using non-smooth PMP based
  techniques. Extensive numerical studies illustrate the three models,
  control techniques and corresponding outcomes. 
\end{abstract}
\end{frontmatter}
\section{Introduction}
\label{sec:intro}

\subsection{Background}
The process of opinion formation and learning by individuals over
social networks has been researched for many decades now. Although
these are, in general, complex processes, determined by the nature of
influences on the individuals over the social network, many simple and
useful mathematical models have been proposed and studied. Most
popular models have the following structure. The social network is
described by a weighted directed graph in which the nodes of the graph
represent the agents or social players and the weighted edges
represent the interactions between the agents. The opinion of an agent
is modeled as a scalar quantity associated with the node corresponding
to that agent. These scalar quantities could represent the strength of
an orientation towards a subject or a topic, in which case the range
of valid values of the scalar could be a subset of the real numbers.
The scalar quantity could also represent a subjective probability
associated with some chance event, in which case the range would be a
subset of $[0,1].$ The opinions at the nodes evolve over time
according to a social interaction model that is described by a
discrete time evolution equation with the opinion at node $i$ at time
$t+1,$ say $x_i(t+1),$ being a function of the opinions of itself and
of its neighbors on the social network at time $t.$ Vector of opinions
at the nodes and continuous time systems have also been considered in
the literature.

The classical opinion evolution, or opinion dynamics, equation is the
French-DeGroot model defined by
\begin{displaymath}
  x_i(t+1) = \sum_{j} A_{i,j} x_j(t)
\end{displaymath}
where $A_{ij}$ is the weight of edge $(i,j).$ A continuous time
version of this model is called the Abelson model and is given by the
following differential equation model.
\begin{displaymath}
  \dot{x}_i(t) = \sum_{j} A_{i,j} \left( x_j(t) - x_i(t) \right)
\end{displaymath}

There are of course several variations of the French-DeGroot
model. The Friedkin-Johnsen model introduces a susceptibility
parameter where the opinion at a node evolves as a convex combination
of the neighbors' opinions and a native prejudice. In the
Hegselmann-Krause model only a set of `trusted' neighbors (defined as
neighbors whose opinions do not differ by more than a threshold) can
influence a node. In Taylor's model, the neighbor's opinions and a set
of static sources, or communication channels, can influence the scalar
opinions at the nodes. A reasonably comprehensive introduction to the
preceding models and their analyses are available in
\cite{proskurnikov2018tutorial-I,proskurnikov2018tutorial-II}. Another
excellent reference is \cite{Acemoglu10}. In this paper we will
consider a variation of the Friedkin-Johnsen model and will be
described in more detail in the next section.

\subsection{Related Work}
The key characteristic of the models that are surveyed in
\cite{proskurnikov2018tutorial-I,proskurnikov2018tutorial-II,Acemoglu10}
is the absence of any influences or inputs external to the social
network. Further, all the nodes behave similarly in the updating of
their opinions. Thus the analysis objective in these early works is in
the asymptotic behavior of the opinions $x_i(t)$ and the convergence
rate to the asymptotic values.

There is now a growing interest in analyzing the effect on opinion
formation in the presence of inputs external to the social network;
see e.g., \cite{Jadbabaie12,Azzimonti18}. The objective of these works
is, like before, to analyze the equilibrium behavior of the opinions
under different models of the external inputs. These studies are
prompted by the need to understand how campaigns on social media
influence the group behavior of a social network. Another variation is
to consider the situation where some of the nodes behave differently
than others; specifically, a small number of the nodes are stubborn
and do not change their opinions while the rest of the nodes change
their opinions according to the usual opinion dynamics model. Example
models and analyses are in
\cite{Acemoglu10a,Acemoglu13,Ghaderi14,Pirani16}.  Once again, the
interest is in the analysis to study the equilibrium behavior.

A key departure from analysis to design towards achieving a desired
objective was, to the best of our knowledge, first considered in
\cite{Borkar11,Borkar15}. They considered choosing the set of stubborn
nodes and their opinion values so as to converge to the predetermined
consensus at the fastest rate subject to cost constraints. A similar
objective for the case of nodes having binary valued opinions is
considered in \cite{Masuda15}. In this paper we consider the design of
external control inputs to achieve a desired control objective on the
opinion values---of achieving a desired opinion profile over the
control horizon. 

We conclude this discussion by pointing out that there are some
superficial similarities between the control of opinions as studied in
this paper and controlled flow of information using epidemic models on
graphs, e.g., \cite{Kandhawy14,Kandhawy16}. We reiterate that they are
not related.  The work in \cite{Eshghi17} has goals most similar to
ours but there are key differences---different objectives and
different constraints. Our interest is in controlling the behaviour
over the entire control period, \cite{Eshghi17} is interested in the
state at the end of the period. For their key result, \cite{Eshghi17}
uses a continuous time model and invokes the Pontryagin maximum
principle (PMP) to show that the control is bang-bang. 

\subsection{Our Contributions}
In the first part of this work, we consider the problem of maintaining
the opinion levels above a predetermined threshold
(\emph{ferment}\footnote{Here, ferment is used to mean ``a state of
  intense activity or agitation. See
  \texttt{https://www.merriam-webster.com/dictionary/ferment}.}) level
at all the network nodes at all times of a finite time horizon, after
allowing for a small startup period. This is to be achieved by control
agents external to the social network that can influence opinion at
some nodes. The control nodes change their opinions according to a
control schedule and are not influenced by other nodes. Equivalently,
some nodes inject `external opinions' into the network. This model is
a natural extension to the goals of \cite{Borkar15,Masuda15}.

The preceding control objective is motivated by campaigns that need to
achieve a minimum interest level, or a `mindshare,' in the population.
This interest level is modeled as a scalar value at the nodes. Another
motivation is the many studies that have reported use of online social
networks to create social ferment to facilitate certain
actions\footnote{For an example news story,
  see\\ https://www.nytimes.com/2018/10/15/technology/myanmar-facebook-genocide.html}. There
have also been news reports about online social networks wanting to
counter such
activities \footnote{https://www.businesstoday.in/buzztop/buzztop-feature/heres-how-whatsapp-plans-to-fight-fake-news-in-india/story/309163.html}. 
%

In the preceding, ferment is to be maintained at minimum cost. In the
second part of the paper, consider the objective of maximizing the
minimum of a scalar function of the opinion vector over a finite time
horizon, subject to a given total cost constraint for the period.

Our contributions and the organization of the rest of the paper are as
below.

In Section~\ref{sec:model} we set up the notation and describe the
opinion dynamics model, a variation of the Friedkin-Johnsen model.  In
this model, in the absence of the control inputs, the opinions at each
node decays to a node-specific quiescent level. We also set up the
optimal control problem of maintaining ferment in the social network
at times $\{T_0, \ldots, T\}$ at minimum cost when nodes that can
receive the control inputs are provided.  In Section~\ref{sec:PMP} we
present a technique to determine an optimal control trajectory based
on the Pontryagin maximum principle (PMP). In
Section~\ref{sec:turnpike} we show that the optimal control trajectory
has the turnpike property, i.e., as the time horizon $T$ becomes
large, the states and the control will be near their equilibrium level
for all but a vanishing fraction of time instants. This property
enables us to pose the choice of the controlled nodes as a
constrained optimization problem. This optimization problem, its
solution under a relaxation, and the possible heuristics are described
in Section~\ref{sec:choosing}. Section~\ref{sec:maxmin} considers the
max-min problem of maximizing the minimum of a scalar function of the
opinion vector over a finite time horizon, subject to a given total
cost constraint and describes the numerical procedure for deriving a
control function for this problem using non-smooth PMP based
techniques. In Section~\ref{sec:sim}, numerical simulations are used
to illustrate the properties of the cost of maintaining ferment on
three different graph models that are widely used to describe social
networks.

\section{Preliminaries}
\label{sec:model}

In this section we first describe the opinion dynamics model, followed
by the control objectives. Toward this, we employ the following
notation. $\mathbb{R}$ denotes the real numbers, $\mathbb{R}_+$ the
positive real numbers, and $\mathbb{N}$ the positive integers. $I_n$
is the $n \times n$ identity matrix. The $i$-th entry of vector $v$ is
denoted by $v_i$. For $a,b \in \mathbb{R}^n$, the relation $a \geq b$
denotes the entry-wise inequality $a_i \geq b_i \text{~for~all~}i \in
\{1,\ldots,n\} $. The transpose of matrix $\mathcal{M}$ is denoted by
$\mathcal{M}^\intercal$. For matrix $\mathcal{B} \in
\mathbb{R}^{n\times m}$, the norm $\|\mathcal{B} \|_0$ denotes the
total number of non-zero entries in $\mathcal{B}$.  The dot product of
vectors $u$ and $v$ is denoted by $\left< u,v \right> $

\subsection{Opinion Dynamics}
A set of $n$ agents, denoted by $\mathcal{N},$ are connected over a
social network. Agent $i$ has a positive scalar-valued opinion,
modeled as the state $x_i(t)\in \mathbb{R}$ at time $t.$ The agents
interact with their neighbors over the social network and evolve their
opinions with time. The social interactions are modeled by the
weighted directed edges of the network graph $\mathcal{G}
=(\mathcal{N},\mathcal{E})$ with the edge set $\mathcal{E}$ modeling
the interaction among the agents. 
The opinion (state) dynamics in the network for each node $i$, in the
absence of external control inputs, is governed by the following
variation of the Friedkin-Johnsen model law.
\begin{equation}
  x_i(t+1) = q_i + \sum_{j:(i,j)\in \mathcal{E}} a_{ij}\left( x_j(t)-
  q_j \right) + a_{ii} \left( x_i(t)-q_i \right)
  \label{eq:free-opinion-dynamics}
\end{equation}
Here $q_i, $ $q_i \geq 0,$ is the quiescent opinion of agent $i$ (it
could be the final state from an exogenous interaction process),
$a_{ij},$ $0 \leq a_{ij} < 1,$ models the strength of the influence of
opinion of agent $j$ on that of agent $i,$ and $a_{ii}$ models the
\emph{stubbornness} of agent $i.$

\begin{assumption} \label{assumption:substochastic}
    We assume $\sum_{j=1}^n a_{ij} < 1$ for all $i,$ i.e., the weight matrix $A=[[a_{ij}]]$ is substochastic.
  \end{assumption}

\begin{remark}
Under this assumption, at any time, the opinion of each
agent has two components---a constant native opinion and an additive
perturbation due to interactions with neighbours. In the absence of
external inputs, all agents regress to their native opinion. In the
context of our motivating scenarios, we believe that this is a more
natural model. 
\end{remark}
The vector form of \eqref{eq:free-opinion-dynamics} can
be written as
\begin{equation}
  x(t+1) = Ax(t)+ (I_n-A)q. 
  \label{eq:free-opinion-dynamics-vector}
\end{equation}
Here $x(t)$ is a column vector with $x_i(t)$ as $i$-th component, $q$
is the column vector of quiescent opinions, and $I_n$ is the identity
matrix. \rev{It can be checked that, without a control input, $x(t) \to
  q.$ } \revv{See that $x(t) = A^{t}x(0) + \sum_{i = 0}^{t}A^i(I_n -A)q$. As $A$ is substochastic, $x(t)$ tends to $q$ as $t$ grows large.}
 Our interest is in maintaining ferment over the finite time horizon
$\{T_0,\ldots,T\},$ through control inputs injected at the
controlled nodes starting at time $t=0.$ 
  \begin{assumption} \label{assumption:structure_of_B}
There are $m$ control inputs, each of which can inject control actions
into exactly one \emph{controlled node.} 
\end{assumption}
Matrix $B \in \{0,1\}^{n
  \times m}$ maps control sources to controlled nodes with
$b_{ij}=1$ if control $j$ is connected to node $i$.
\begin{remark}
\rev{We also remark that $b_{ij}$ could in general be any positive real number. Making it take values in $\{0,1\}$ allows us
  to provide additional results.}
\end{remark}

%
%
In the presence of the control inputs, the opinion dynamics take the
following form.
\revv{\begin{equation}
  x(t+1) = Ax(t) + Bu(t) + (I_n-A)q ~~~ \text{for~} t = 0,\ldots,T-1
  \label{eq:dynamics}
\end{equation}}
Here $u(t) \in \mathbb{R}^m$ is the column vector of control actions
applied at time instant $t.$ In the rest of the paper, for brevity, we
denote the state-action trajectory $(x(t), u(t))_{t=0}^{T-1}$ by
$(x,u)$.

\rev{
  \begin{assumption} \label{assumption:controllable}
    We assume that the $(A,B)$ is controllable.
  \end{assumption}

  Assumption~\ref{assumption:controllable} is basic in this context~\cite{Eshghi17}. 
  
  \begin{assumption} \label{assumption:convex-cost}
    We assume that the control actions have a convex cost.
  \end{assumption}
  
  Assumption~\ref{assumption:convex-cost} is reasonable because of the law of diminishing returns;
  in this context this means that effecting marginal change becomes
  harder with increasing values. To anchor the discussion, we define
  the following cost function on the control inputs
  \begin{equation}
    \mathbb{R}^m \ni \mu \mapsto c(\mu):= \mu^\intercal R \mu \in
    \mathbb{R}. 
    \label{eq:cost}
  \end{equation}
  Here $R \in \mathbb{R}^{m \times m}$ is symmetric and positive
  definite.
}

\subsection{Control Objectives}

We view the problem from the point of view of the external agent
applying the control inputs and consider the following three
different, yet related, objectives---total ferment~\eqref{eq:TF-OCP},
group ferment~\eqref{eq:GF-OCP}, and maxmin group
ferment~\eqref{eq:MF-OCP}. These are detailed below.

In problem~\eqref{eq:TF-OCP}, the objective is to have the opinion at
every node exceed a specific threshold for $T_0 \leq t \leq T.$
Specifically, let $\tau_i$ be the minimum opinion level that needs to
be maintained at node $i$ in the time interval $\{T_0,\ldots,T\},$
i.e., $x_i(t) \geq \tau_i,$ for all $i$ and for all $t \in
\{T_0,\ldots,T\}.$ Let $\tau = [\tau_i]$ be the column vector of the
thresholds. This imposes a state constraint given by
\begin{equation}
  x(t) \in \begin{cases}
    \mathbb{R}^n &\mbox{for~} t = 0,\ldots, T_0-1,  \\
    \mathbb{X}_{TF}   &\mbox{for~} t = T_0,\ldots, T ,
  \end{cases}
  \label{eq:total-ferment}
\end{equation}
where $\mathbb{X}_{TF} = \{y \in \mathbb{R}^n~ | ~y\geq \tau\}.$

The problem of minimizing the cost of~\eqref{eq:TF-OCP} is stated as
an optimal control problem in the following Lagrange form.
\begin{align} 
 \tag{TF} \label{eq:TF-OCP}
  \begin{split}
    &\underset{u}{\text{minimize}}~~ J_{TF}(\bar{x},u) = \sum_{t=0}^{T-1}
    c\left( u(t) \right) \\ 
    &\text{subject~to}~~ 
    \begin{cases}      
      \text{state~dynamics~of~\eqref{eq:dynamics}},\\ 
      \text{state~constraints~of~\eqref{eq:total-ferment}} \\ 
      x(0) = \bar{x} \in \mathbb{R}^n \text{~(given).}
    \end{cases}
  \end{split}
\end{align}
Problem~\eqref{eq:TF-OCP} is a constrained LQ optimal control
problem. The cost of an optimal control-action trajectory
for~\eqref{eq:TF-OCP} with initial state $\bar{x}$ will be denoted by
$J^*_{TF}(\bar{x})$, i.e.,
\begin{equation} 
  \label{eq:optimal-cost-total}
  J^*_{TF}(\bar{x}) := \underset{u}{~ \inf ~} J_{TF}(\bar{x},u).   
\end{equation}

For problem~\eqref{eq:GF-OCP} we first define the following function
of the opinions. Let $\psi(x(t)):\mathbb{R}^n \rightarrow \mathbb{R}$
be an \rev{elementwise} increasing and differentiable function defined
on the opinion vector $x.$ For~\eqref{eq:GF-OCP} we will require that
$\psi(x(t)) \geq k n$ for $T_0 \leq t \leq T$ and some $0 < k <
1$. Define the set $\mathbb{X}_{GF} := \{y \in \mathbb{R}^n~ |
~\psi(y) \geq kn \}.$ The state constraints for~\eqref{eq:GF-OCP} are
given by
\begin{equation}
    x(t) \in \begin{cases}
      \mathbb{R}^n &\mbox{for~} t = 0,\ldots, T_0-1,  \\
      \mathbb{X}_{GF}   &\mbox{for~} t = T_0,\ldots, T .
  \end{cases}
  \label{eq:group-ferment}
\end{equation}

\rev{For $\psi(\cdot)$ any of the functions that are used in
  classification problems could be used. In this paper, we will use
  the following form of $\psi$, chosen for the convenience in
  computing its derivative: for $a>0$ and $\tau>0,$ }
\begin{displaymath}
  \mathbb{R}^n \ni y \mapsto \psi(y) = \sum_{i=1}^n \frac{1}{1+e^{-a(y_i-\tau)}}.
\end{displaymath}
With a suitable choice of $a$, the requirement that $\psi(x(t)) > kn$
for $T_0 \leq t \leq T$, is a soft proxy for the requirement that the
fraction of nodes with opinions larger $\tau$ be above $k.$ The
constrained LQ optimal control problem to achieve~\eqref{eq:GF-OCP}
will be as under.
\begin{align} 
 \tag{GF} \label{eq:GF-OCP}
  \begin{split}
    &\underset{u}{\text{minimize}}~~ J_{GF}(\bar{x},u) = \sum_{t=0}^{T-1}
    c\left( u(t) \right) \\ 
    &\text{subject~to}~~ 
    \begin{cases}      
      \text{state~dynamics~\eqref{eq:dynamics}},\\ 
      \text{state~constraints~\eqref{eq:group-ferment}} \\ 
      x(0) = \bar{x} \in \mathbb{R}^n \text{~(given).}
    \end{cases}
  \end{split}
\end{align}
The cost of an optimal control-action trajectory for~\eqref{eq:GF-OCP}
with initial state $\bar{x}$ will be
\begin{equation} 
  \label{eq:optimal-cost-total-GF}
  J^*_{GF}(\bar{x}) := \underset{u}{~ \inf ~} J_{GF}(\bar{x},u).   
\end{equation}

The preceding two optimal control problems are minimum cost
formulations. An alternative would be to have a total cost
constraint. For such a situation we define a maxmin optimization
problem as follows. We will continue to use the function $\psi$
defined above and consider the following optimal control problem.

\begin{align} 
\tag{MF}  \label{eq:MF-OCP}
  \begin{split}
    &\underset{u}{\text{maximize}}~~ \underset{t}{\text{minimum}}
    ~~ \psi(x(t)) \\
    &\text{subject~to}~~ 
    \begin{cases}
      \text{state~dynamics~\eqref{eq:dynamics}},\\ 
      J_{MF}(\bar{x},u) = \sum_{t=0}^{T-1} c\left( u(t) \right) \ \leq \ C\\
      x(0) = \bar{x} \in \mathbb{R}^n \text{~(given).}
    \end{cases}
  \end{split}
\end{align}

In the next section we describe a method to obtain an optimal control
trajectory for problems \eqref{eq:TF-OCP} and \eqref{eq:GF-OCP} based
on the Pontryagin maximum principle. The optimal control trajectory
for problem \eqref{eq:MF-OCP} will be described in
Section~\ref{sec:maxmin}.

\section{Optimal Control via Pontryagin Maximum Principle}
\label{sec:PMP}

In this section we give a proof that an optimal solution exists for the optimal control problems~\eqref{eq:TF-OCP} and~\eqref{eq:GF-OCP}. This result, given formally in Theorem~\eqref{thm1}, holds under Assumption~\eqref{assumption:controllable}. We show that it also holds under the following weaker assumption on the pair $(A,B):$
\revv{
\begin{assumption} \label{assumption:path_from_controllable}
 For every node in the network, there exists a directed path to it from at least one of the controlled nodes.
\end{assumption}
}
Let $d_i$ be the number of hops from the nearest controlled node to node $i$ and and let $d_{\max} = \max_i d_i$. Then we have the following result.

\begin{thm} \label{thm1}
  Consider problems~\eqref{eq:TF-OCP} and~\eqref{eq:GF-OCP} and suppose that Assumption~\ref{assumption:path_from_controllable} holds. For $T_0 > d_{\max}$, an optimal solution to the
  problems~\eqref{eq:TF-OCP} and~\eqref{eq:GF-OCP} exists.
\end{thm}

\begin{pf}
\revv{ 
 We first show that the problem is feasible. Note that the feasibility of problem~\eqref{eq:TF-OCP} implies the feasibility of problem~\eqref{eq:GF-OCP}. This is true because for all parameter $k$ and function $\psi(\cdot)$ defining set $\mathbb{X}_{GF}$, there is a vector $\tau$ defining set $\mathbb{X}_{TF}$ such that $y \in \mathbb{X}_{TF} $ implies $ y \in \mathbb{X}_{GF}.$ In other words, the state constraints of problem~\eqref{eq:GF-OCP} are satisfied if the state constraints of problem~\eqref{eq:TF-OCP} are satisfied for some appropriately chosen threshold $\tau.$
 Recall that the dynamics of the system are as follows:
 \begin{align}
x(1) &= Ax(0) + Bu(0) + (I_n -A)q \\
x(t) &= A^n x(0) + (A^{t-1}B \ldots AB ~B)  \label{eq:xt}
      \begin{pmatrix}
          u(0)\\           
           \vdots \\
           u(t-1)
    \end{pmatrix} \\
    & + (A^{t-1} \ldots A ~I) 
      \begin{pmatrix}
         (I_n -A)q \\           
           \vdots \\
         (I_n -A)q
    \end{pmatrix} \nonumber
 \end{align}
 Note that all entries of $A$ and $B$ are non-negative. Recall from Assumption~\ref{assumption:structure_of_B} that the matrix $B$ maps a control input to exactly one controlled node. This implies that $B$ has a $1$ entry in its $i$-th row if the $i$-th node is a controlled node. See that $A^tB$ is a set of $m$ columns, one corresponding to each controlled node. The $l$-th column of $A^tB$ contains the weights of the $t$-length path from the $l$-th controlled node to all nodes of the network. From Assumption~\ref{assumption:path_from_controllable}, there exists a directed path of length at most $d_{\max}$ from at least one of the controlled nodes to all nodes of the network. Therefore for $T_0 > d_{\max},$ the matrix $(A^{T_0-1}B \ldots AB ~B)$ has no row with all zeros. Here, the $i$-th row of $(A^{T_0-1}B \ldots AB ~B)$ being zero implies that node $i$ cannot be influenced in $T_0$ time steps with the given set of controlled nodes. Let for the $j$-th row of $(A^{T_0-1}B \ldots AB ~B)$, the $i$-th column has one of the non-zero entries. Then, see from~\eqref{eq:xt} that it is possible to apply a large enough control on the $i$-th entry of $(u(0), \cdots, u(T_0-1))$ to achieve $x_j(t) \geq \tau_j.$ This is true for all $j \in \{1, \ldots, n\}.$ Therefore, for $T_0 >  d_{\max}$, there exists a control action trajectory $(u(0), \cdots, u(T_0-1))$ for which  $x(T_0) \in  \mathbb{X}_{TF}$. Note that we do not require $(A^{T_0-1}B \ldots AB ~B)$ to be of full rank to ensure $x(T_0) \in  \mathbb{X}_{TF}$ because we do not want to take $x(T_0)$ to a specific point. Rather, we require it to be larger than a threshold, which can be ensured by applying a large enough control if $(A^{T_0-1}B \ldots AB ~B)$ doesn't have a zero row. See that $(A^{T_0-1}B \ldots AB ~B)$ has no negative entries because $A$ and $B$ have no negative entries. Now, we repeat this for all time steps $t \in \{T_0 +1, T_0+2, \cdots, T\}$. The sum of the control action trajectories obtained on solving these $T-T_0 +1$ problems is an example of a feasible solution for the problem~\eqref{eq:TF-OCP}.
  
Now consider the problem as an optimization problem over $(u(0), \cdots, u(T-1)).$ Since the problem is feasible, there exists one trajectory $((\bar u)(0), \cdots, (\bar u)(T-1))$ that satisfies all constraints. Pick a compact set $K$ such that outside $K$ the cost is larger than the cost for $((\bar u)(0), \cdots, (\bar u)(T-1))$. Such a compact $K$ exists due to the near-monotonicity\footnote{\rev{$f:
    \Re^n \to \Re$ is near monotone if $\lim_{r \to \infty} \inf_{y
     \notin B(0,r)} f(y) = + \infty.$}} of the cost $c(\cdot)$. The cost of $(u(0), \cdots, u(T-1))$ is continuous and $K$ is compact. Weierstrass's theorem~\cite{royden1968real} guarantees the existence of a minimum.  }
\end{pf}

In the rest of the paper we will assume $T \gg T_0 \geq d_{\max}.$

We use the necessary conditions of optimality from  the discrete
time PMP \cite{paruchuri2019discrete} to solve the
problems~\eqref{eq:TF-OCP} and~\eqref{eq:GF-OCP}. The costate variable
is  $\left(\lambda(t)\right)_{t=0}^T \in \mathbb{R}^n$. Define
the Hamiltonian
\begin{align}
  &\mathbb{R}\times \mathbb{R}^n \times \mathbb{N} \times \mathbb{R}^n
  \times \mathbb{R}^m \ni (\nu,\zeta,\xi,\mu) \mapsto
  H(\nu,\zeta,\xi,\mu) \nonumber \\
  &:= \zeta^\intercal \left( A\xi + B\mu+(I_n-A)q \right) -\nu c\left(
  \mu \right) \in \mathbb{R}.
\end{align}

Let $(x^*(t))_{t=0}^{T}$ and $(u^*(t))_{t=0}^{T-1}$ be an optimal
state-action trajectory that solves~\eqref{eq:TF-OCP}
(similarly~\eqref{eq:GF-OCP}). PMP asserts that there exists $\nu \in
\{0,1\}$, a sequence $(\lambda(t))_{t=0}^T$, and a sequence
$(\eta(t))_{t=T_0}^{T-1}$ such that
\begin{align}
  \lambda(T) & = 0 \nonumber \\
  \lambda(t-1) & = \frac{\partial}{\partial \xi}
  H\left(\nu,\lambda(t),x^*(t), u^*(t)\right) - \eta(t) 
  \label{eq:lambda-of-T}
\end{align}
for $t = 1,\ldots,T-1$, such that the scalar $\nu$ and the sequence
$(\lambda(t))_{t=0}^T$ do not identically vanish. The Hamiltonian
maximization condition asserts that an optimal control trajectory
$(u^*(t))_{t=0}^{T-1}$ must satisfy
\revv{\begin{align}
 \left. \frac{d}{d \mu}\right\vert_{\mu =  u^*(t)}\nu c \left( \mu \right) = B^\intercal \lambda(t)
  \nonumber 
\end{align} }
for $t = 0, \ldots,T-1$. For our quadratic cost function,
\begin{align}
  u^*(t) = \frac{1}{2\nu} R^{-1} B^\intercal\lambda(t).
\end{align}
\revv{From PMP, $\nu$ either takes value $0$ or $1$. The
  case of $\nu = 0$ corresponds to abnormal control~(\cite{clarke2013functional}, Theorem 9.1 on page 179).}

Multiplier $\eta(t) \in \mathbb{R}^n$ takes only non-negative
values by definition. Further, from the complementary slackness condition 
\begin{align}
  \nonumber \eta_i(t)(x_i(t)-\tau_i) = 0 
\end{align}
for $i = 1,\ldots,n,$ and $t = T_0,\ldots,T,$ where $\tau$ is the
required ferment level.

The PMP provides necessary conditions for optimality in the form of a
well-posed system of boundary value problems. There are several
computational procedures that may be used to obtain the optimal
control sequences, including shooting methods and homotopy based
methods; see, e.g.,~\cite{trelat2012optimal} for a detailed
survey. Since most of these techniques are well known, we will not
elaborate on them any more. Instead, we will explore the interesting
turnpike phenomenon that the solutions to~\eqref{eq:TF-OCP}
and~\eqref{eq:GF-OCP} exhibit. This means that as time
horizon $T$ becomes large, the optimal state-action trajectories stay
close to certain equilibrium pairs for all but a vanishing fraction of
time instants. This is investigated in the next section. We will also
see that this behavior allows us to simplify the optimal control
problem and provides mechanisms to choose most efficient, and
effective controlled nodes.

\section{Turnpike Behavior of the Optimal Control}
\label{sec:turnpike}

In this section we demonstrate that under a mild condition on $T_0$,
the optimal state-action trajectories for~\eqref{eq:TF-OCP}
and~\eqref{eq:GF-OCP} possess a turnpike property. Recall that the
turnpike property implies that as the time horizon $T$ becomes large,
the state and control remain close to an equilibrium point for all but
a vanishing fraction of time instants for a discrete time optimal
control problem. It was first observed and studied by von Neumann
\cite{von1971model} and Dorfman et al., \cite{dorfman1987linear} in
the context of optimal control in economics. The turnpike property is
closely related to the system theoretic properties of
\textit{dissipativity} and \textit{strict
  dissipativity}. Dissipativity formalizes the condition that a system
cannot store more energy than supplied from the outside; strict
dissipativity requires, in addition, that some energy is dissipated to
the environment. The relation between (strict) dissipativity and the
turnpike property was first established in \cite{grune2013economic}
and later generalized in \cite{grune2016relation}. Specialized results
for constrained discrete time linear quadratic optimal control
problems are in \cite{grune2018turnpike}. We begin with the
following definitions.
\begin{definition} (\hspace{1sp}\cite{grune2013economic})
  A state-action pair
  $( x^e, u^e ) \in \mathbb{R}^n \times \mathbb{R}^m$ 
  is called an equilibrium of~\eqref{eq:dynamics} if
  $f( x^e, u^e ) = x^e$, and $(x^e, u^e)$
  satisfies the given state and action constraints.
\end{definition}
For the dynamics in~\eqref{eq:dynamics} and 
constraints in~\eqref{eq:total-ferment}, the equilibrium point
for~\eqref{eq:TF-OCP} can be explicitly found by solving the following
convex problem. 
\begin{align} \label{eq:equilibrium}
  \begin{split}
    &\underset{u^e}{\text{minimize}} ~~   (u^e)^\intercal R u^e \ \ , \\
    &\text{subject to} 
    \begin{cases}
      x^e = Ax^e + Bu^e + (I_n-A)q \ ,  \\
      x^e \geq \tau \ . \\
    \end{cases}
  \end{split}
\end{align}

The equilibrium for~\eqref{eq:GF-OCP} can be analogously defined. We have the following lemma (see
  Appendix~\ref{app:lemma3-proof} for proof) for the existence of such an equilibrium.
  \begin{lemma}
   Under Assumption~\ref{assumption:path_from_controllable}, an equilibrium point, as defined in~\eqref{eq:equilibrium}, exists.
    \label{lemma:lem3}
  \end{lemma}

\begin{definition} (\hspace{1sp}\cite[Definition 2.4]{grune2018turnpike})
  Let $\mathcal{K}$ be the set of continuous and strictly increasing
  functions, i.e., $\mathcal{K} := \{ h : \mathbb{R}_+ \to
  \mathbb{R}_+ |h~\text{is strictly increasing, continuous, \& } h(0)
  = 0\}.$
  Given an equilibrium $( x^e, u^e )$ of~\eqref{eq:dynamics}, the
  system~\eqref{eq:TF-OCP} (resp.~\eqref{eq:GF-OCP}) is called
  \emph{strictly dissipative} with respect to supply rate
  $\mathbb{R}^m \ni u \mapsto c(u) - c(u^e) \in \mathbb{R}$ if there
  exists a function $g : \mathbb{X} \to \mathbb{R}$ bounded from below
  and a function $\rho \in \mathcal{K}$ such that for all $x \in
  \mathbb{X}$ and $u \in \mathbb{R}^m$ satisfying $f(x,u) \in
  \mathbb{X}$, we have
  \begin{align}
    c(u) - c(u^e) + g(x) - g(f(x, u)) \geq \rho(\| x - x^e \|).
  \end{align}
  The system~\eqref{eq:TF-OCP} (resp.~\eqref{eq:GF-OCP}) is called
  \emph{dissipative} if the preceding property holds with $\rho \equiv
  0$.
\end{definition}
A readily verifiable and sufficient condition for the
system~\eqref{eq:TF-OCP} (resp.~\eqref{eq:GF-OCP}) to be strictly
dissipative can be obtained from
(\hspace{1sp}\cite{grune2018turnpike}, Lemma 4.1) rewritten as below.
\begin{lemma}
  Let $c(\cdot)$ be as defined in~\eqref{eq:cost}. Given a positive
  definite matrix $P \in \mathbb{R}^{n \times n}$, there exists $q \in
  \mathbb{R}^n$ such that the constrained LQ system~\eqref{eq:TF-OCP}
  or~\eqref{eq:GF-OCP} is strictly dissipative with storage function
  $g(x) = x^\intercal P x + q^\intercal x$ if and only if the matrix
  $P - A^\intercal P A$ is positive definite.
\end{lemma}

To prove that system~\eqref{eq:TF-OCP} (resp. \eqref{eq:GF-OCP}) is
strictly dissipative, we need to show that with $P=I_n$, $I_n -
A^\intercal A$ is positive definite. However, since $A$ is
sub-stochastic by definition, it follows that~\eqref{eq:TF-OCP}
(resp.~\eqref{eq:GF-OCP}) is strictly dissipative.

We now define two variants of the turnpike property in relation
to~\eqref{eq:TF-OCP} and~\eqref{eq:GF-OCP}, adapted from
\cite{grune2016relation, grune2018turnpike}. Denote the state
trajectory under the action sequence $u$ by
$(x_u(t,\bar{x}))_{t=0}^{T}$, where $\bar{x}$ is the initial
state. Moreover, denote the total cost of the state-action trajectory,
$((x_u(t,\bar{x}))_{t=0}^{T},u)$ by $J_{TF}(\bar{x},u)$ as
in~\eqref{eq:TF-OCP} or by $J_{GF}(\bar{x},u)$ as
in~\eqref{eq:GF-OCP}.
\begin{definition}  \label{def:ne_turnpike}
  The optimal control problems~\eqref{eq:TF-OCP} and~\eqref{eq:GF-OCP}
  have the \emph{near-equilibrium turnpike property} at an equilibrium
  $(x^e, u^e)$ if for each $\rho > 0$, $\epsilon > 0$, and $\delta >
  0$ there exists a constant $C_{\rho,\epsilon,\delta}>0$ such that
  for all $\bar{x} \in \mathbb{R}^n$ with $\|\bar{x}-x^e\|\leq \rho$,
  all $T \in \mathbb{N}$, and all trajectories
  $(x_u(t,\bar{x}))_{t=0}^{T}$ satisfying
  \begin{align} \label{netp:1}
    J_{TF}(\bar{x},u) \leq T c(u^e) + \delta \\
    J_{GF}(\bar{x},u) \leq T c(u^e) + \delta
  \end{align}
  for some admissible $u$, with 
  \begin{align} \label{netp:2}
    \# \{ t \in \{T_0,\ldots,T\} ~|~ \|x_u(t,\bar{x}) - x^e \|~ >
    \epsilon \} \leq C_{\rho,\epsilon,\delta }.
  \end{align}
  
\end{definition}

\begin{definition} \label{def:turnpike}
  The optimal control problem given by~\eqref{eq:TF-OCP}
  (resp.~\eqref{eq:GF-OCP}) has the \emph{turnpike property} at an
  equilibrium $(x^e, u^e) $ of~\eqref{eq:dynamics} on a set
  $\mathbb{X}_{tp} \subset \mathbb{R}^n$ if for each compact set $K
  \subset \mathbb{X}_{tp} $ and for each $\epsilon > 0$ there exists a
  constant $C_{K,\epsilon}>0$ such that for all $\bar{x} \in K$, all
  $T \in \mathbb{N}$, the optimal state trajectories
  $(x^*(t))_{t=0}^T$ with initial value $\bar{x}$ satisfy
  \begin{align} 
    \# \{ t \in \{T_0,\ldots,T\} ~|~ \|x^*(t) - x^e \|~ > \epsilon \}
    \leq C_{K,\epsilon}.
  \end{align}  
\end{definition}

\begin{definition}\label{def:cheaply}(\hspace{1sp}\cite{grune2016relation})
  The equilibrium $x^e$ is cheaply reachable for system~
  \eqref{eq:TF-OCP} (resp. system~\eqref{eq:GF-OCP}) if there exists a
  constant $D > 0$ with $J^*_{TF}(\bar{x}) \leq T c(u^e) + D$
  (resp.$J^*_{GF}(\bar{x}) \leq T c(u^e) + D$) for all $\bar{x} \in
  \mathbb{R}^n$ and all $T \in \mathbb{N}$.
\end{definition}

In the preceding, $\#S$ denotes the number of elements of a finite set
$S$. These definitions imply the following. If a control problem
possesses the near-equilibrium turnpike property of (Definition
\ref{def:ne_turnpike}), then the trajectories for which the associated
cost is close to the steady state value stay in a neighborhood of
$x^e$ for most of the time.
If a control problem possesses the turnpike property (Definition
\ref{def:turnpike}), then the optimal state trajectories stay in a
neighborhood of $x^e$ most of the time. They can be far from $x^e$ for
at most a bounded number of time instants, the bound being independent
of $T.$ We assume that $T$ is far from this bound. In fact, we will
see from our simulations that the trajectories approach $x^e$ very
quickly.

From \cite[Theorem 5.3]{grune2013economic} we know that strict
dissipativity implies near-equilibrium turnpike property. Since we
have shown that systems~\eqref{eq:TF-OCP} and~\eqref{eq:GF-OCP}
are strictly dissipative, they also have the near-equilibrium turnpike
property. However, we need the turnpike property as defined in
Definition \ref{def:turnpike} to show that the solution obtained using
the PMP, i.e., an optimal state-action trajectory, stays near the
equilibrium. To this end, we invoke \cite[Lemma
  3.9(b)]{grune2016relation} that is adapted as follows. 

\begin{lemma} \label{lem:cheaply}
  If the optimal control problems~\eqref{eq:TF-OCP} and~\eqref{eq:GF-OCP}
  exhibit the near-equilibrium turnpike property at $(x^e, u^e)$ and
  $x^e$ is cheaply reachable as defined in~\ref{def:cheaply}, then
  they also have the turnpike property at $(x^e, u^e)$.
\end{lemma}

For problems~\eqref{eq:TF-OCP} and~\eqref{eq:GF-OCP}, the state
constraints, $x \in \mathbb{X}_{TF}$ and $x \in \mathbb{X}_{GF}$
respectively are not active for $t \in \{0,\ldots, T_0-1\}$. We use
this property to prove that for $T_0$ large enough, $x^e$ is cheaply reachable as part of
Theorem~\ref{thm:turnpike} below, which is the main result. 

\begin{thm} \label{thm:turnpike}
  For $T_0$ large enough, the optimal control problems~\eqref{eq:TF-OCP} and~\eqref{eq:GF-OCP}
  have the turnpike property.
\end{thm}

\begin{pf}
  From Lemma~\ref{lem:cheaply}, we know that if $x^e$ is cheaply
  reachable from all possible initial states $\bar{x}$, then the proof
  is complete. Consider the control action trajectory $\tilde{u}$ that
  steers the state from $x(0)=\bar{x}$ to $x(T_0)=x^e$ with
  minimum cost, and thereafter it maintains the state at $x^e$ by
  applying action $u^e$. Here the parameter $T_0$ is picked as the
  smallest integer such that $x(T_0) = x^e$ is achievable for all
  $\bar{x} \in \mathbb{R}^n$. For large enough $T$ and a full rank
  Gramian matrix of the system, such a $T_0$ is guaranteed to exist
  for all initial states $\bar{x}$. This follows from the fact that
  the system with a full-rank Gramian is controllable and no
  state-action constraints are imposed on the system for $t \leq
  T_0$. The cost of $\tilde{u}$ is: $J_{TF}(\bar{x},\tilde{u}) =
  (T-T_0) c(u^e) + D^*(\bar{x})$, where $D^*(\bar{x})$ is the finite
  cost of steering the system from $x(0) = \bar{x}$ to $x(T_0) = x^e$
  in minimum cost.  By definition, no control trajectory can have a
  cost smaller than $J^*_{TF}(\bar{x})$ as defined
  in~\eqref{eq:optimal-cost-total}, which means:
  \begin{align}
    J^*_{TF}(\bar{x}) &\leq J_{TF}(\bar{x},\tilde{u}) = (T-T_0) c(u^e)
    + D^*(\bar{x}). \\
    \intertext{Which implies that} J^*_{TF}(\bar{x}) &\leq T c(u^e) +
    D^*(\bar{x}). \label{proof_step}
  \end{align}
  From the definition of cheap
  reachability~\eqref{def:cheaply},~\eqref{proof_step} ensures that
  $x^e$ is cheaply reachable and finishes the proof
  for~\eqref{eq:TF-OCP}. The proof for~\eqref{eq:GF-OCP} follows
  similarly.
\end{pf}
\begin{remark}
\rev{We remark that in the preceding proof, we did not need $b_{ij}
  \in \{0,1\}$ and hence the theorem is valid for $ b_{ij} \in
  \mathbb{R}.$ However, the next section will require that $b_{ij}\in
  \{0,1\}.$ }
  \end{remark}

\begin{figure}[htb]
  \centering {\includegraphics[width=0.9\columnwidth,
      height=0.6\columnwidth]{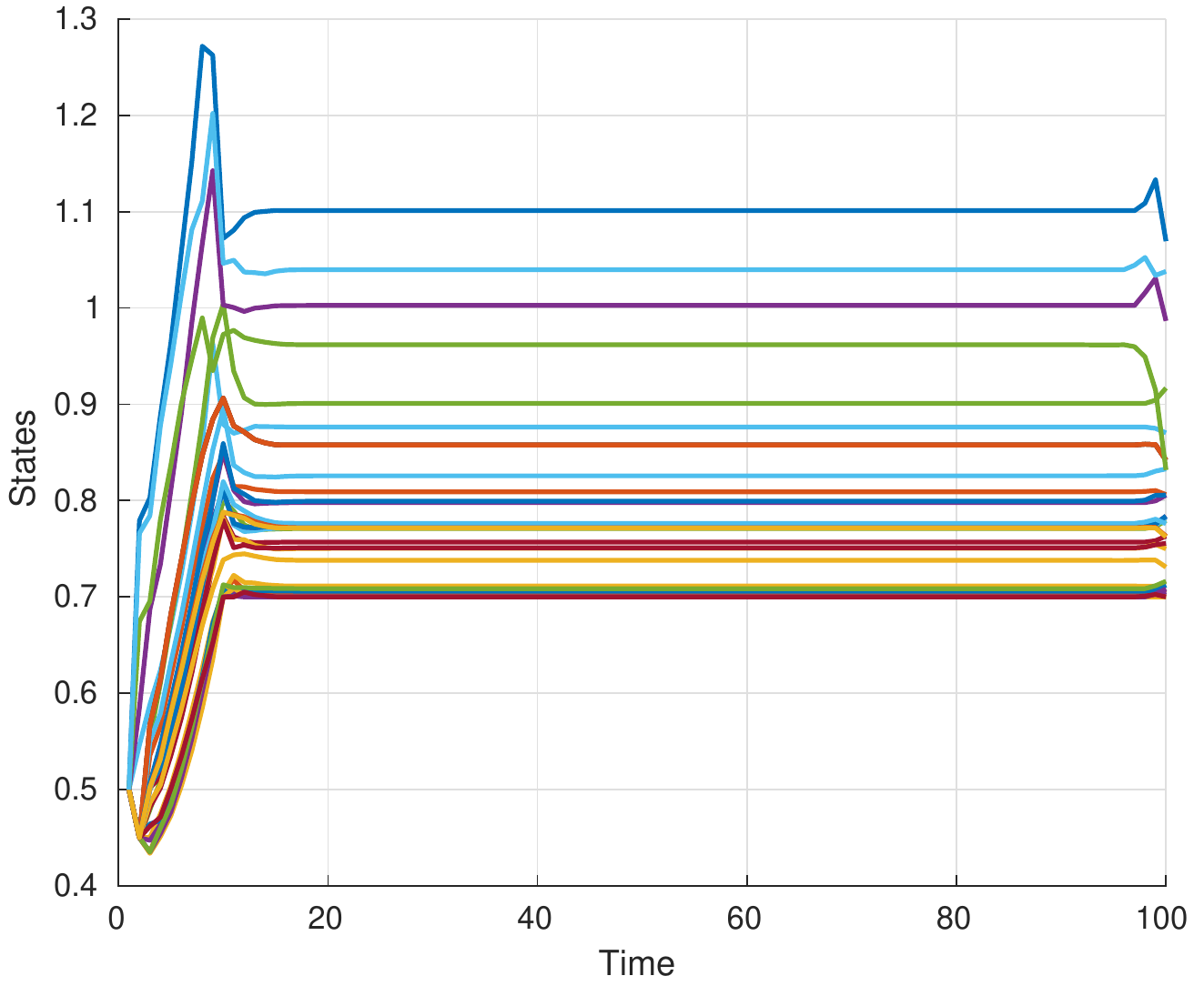}}
  \caption{A typical state trajectory for our optimal control
    problem. Each curve depicts the state trajectory corresponding to
    a node of the underlying graph. The experiment was done on the
    Zachary's Karate Club friendship network, with $n = 34$ nodes, $T
    = 100$ time step, and $T_0 = 10$. The initial opinion level is
    $0.5$ on all nodes and the threshold level is $\tau = 0.7$. There
    are $m = 5$ controlled nodes. Observe that the equilibrium
    values are reached shortly after $t = T_0$. This is evident from
    the fact that $\|x(12)-x^e\|_2 / \|x^e\|_2 = 0.009$ and
    $\|x(15)-x^e\|_2 / \|x^e\|_2 = 0.00076$.}
  \label{fig:t1}
\end{figure}

\begin{figure}[htb]
  \centering
  {\includegraphics[width=0.9\columnwidth,
      height=0.6\columnwidth]{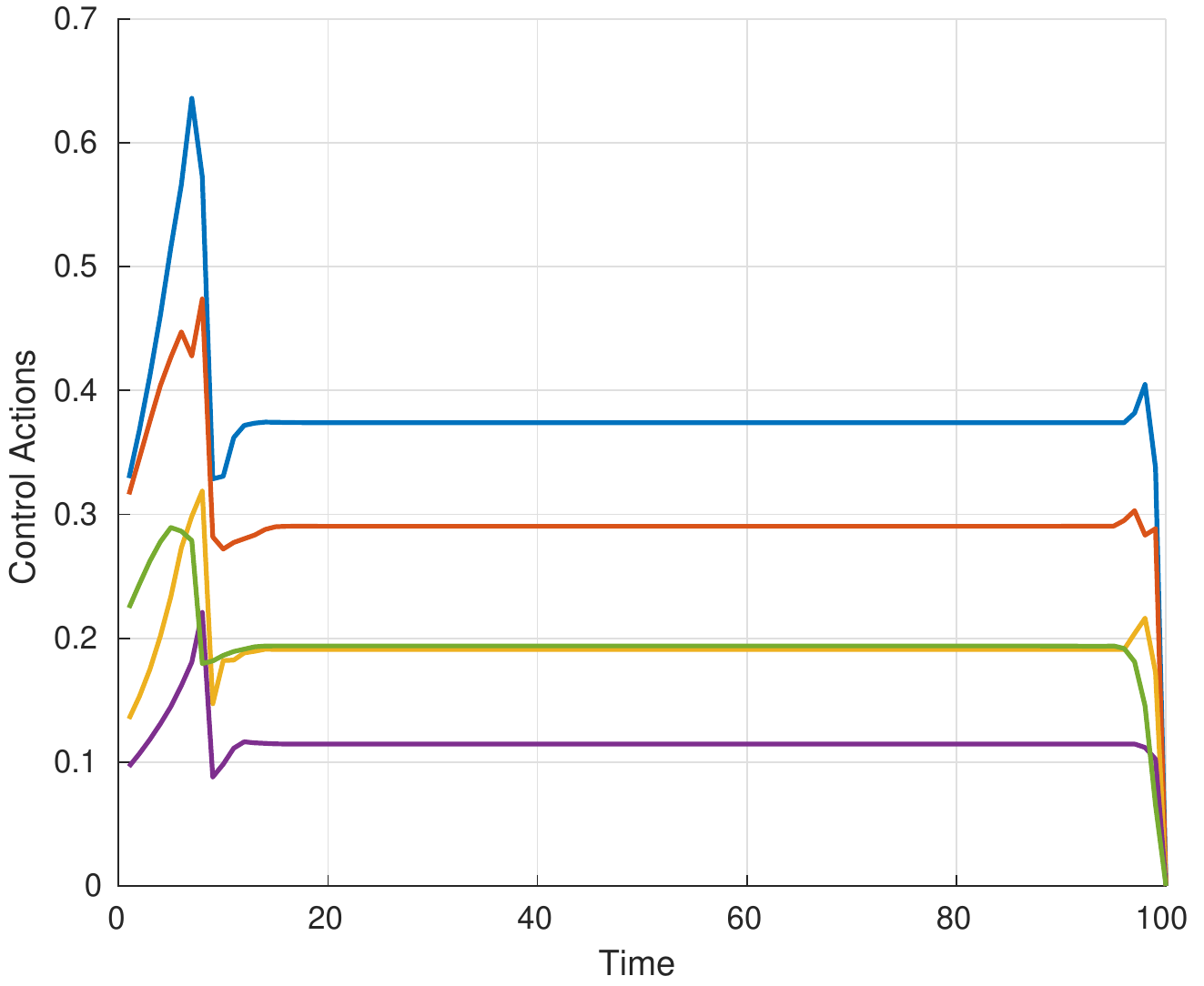}}
  \caption{A typical control action trajectory for our optimal control
    problem. Each curve corresponds to the control actions at one of
    $m = 5$ controlled nodes. The experimental setup is same as for
    Fig.~\ref{fig:t1}. It can be observed that the equilibrium values
    are reached shortly after $t = T_0$.This is evident from the fact
    that $\|u(12)-u^e\|_2 / \|u^e\|_2 = 0.020$ and $\|u(15)-u^e\|_2 /
    \|u^e\|_2 = 0.00089$.}
  \label{fig:t2}
\end{figure}

Figs.~\ref{fig:t1} and~\ref{fig:t2} show a typical turnpike state and
control action trajectories for one instance of problem
\eqref{eq:TF-OCP} on the Zachary's Karate Club friendship network (see
\cite{Zachary1977} for details of the origin of this network) with $n
= 34$ nodes, $T = 100$ time steps and $T_0 = 10$. The initial opinions
are $\bar{x}_i = 0.5$ for all $i \in [1, \ldots, 34]$. The threshold
is $\tau = 0.7$ and the quiescent level is $q=0$ implying that the
interest in the topic dies out eventually in the absence of external
control. There are $m=5$ controlled nodes. It can be seen from the
plots in Figs.~\ref{fig:t1} and~\ref{fig:t2} that the state-action
trajectories stay at the equilibrium for most time instants after
$t=T_0$.

\begin{figure}[htb]
  \centering {\includegraphics[width=0.9\columnwidth,
      height=0.6\columnwidth]{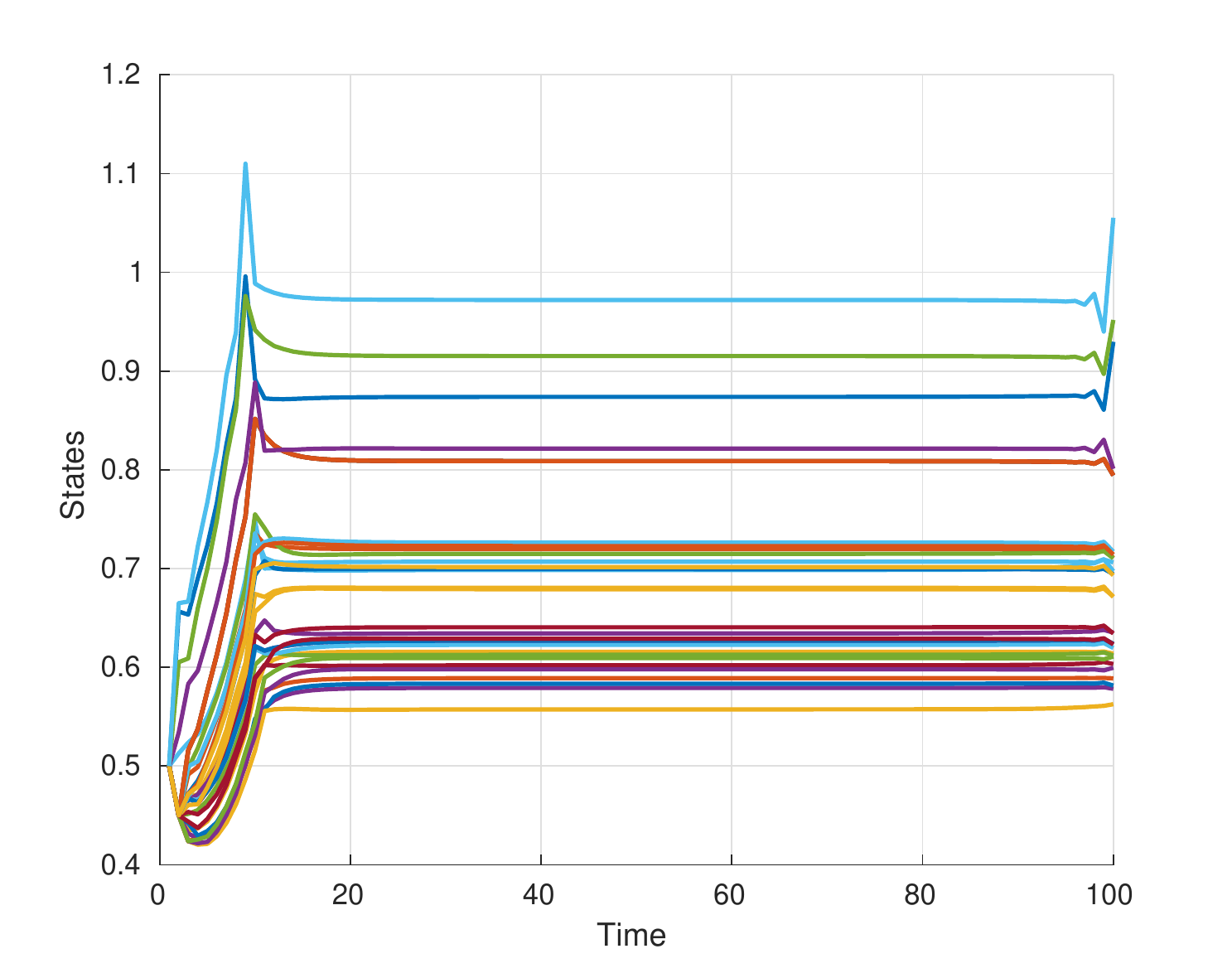}}
  \caption{\rev{A typical state trajectory for the Group Ferment
      optimal control problem. Each curve depicts the state trajectory
      corresponding to a node of the underlying graph. The experiment
      was done on the Zachary's Karate Club friendship network, with
      $n = 34$ nodes, $T = 100$ time step, and $T_0 = 10$. The initial
      opinion level is $0.5$ on all nodes and the threshold level is
      $\tau = 0.7$. There are $m = 5$ controlled nodes. The values
      of parameters $k$ and $a$ are $0.5$ and $1$ respectively.}}
    \label{fig:grp_typical_x}
\end{figure}

\begin{figure}[htb]
  \centering
  {\includegraphics[width=0.9\columnwidth,
      height=0.6\columnwidth]{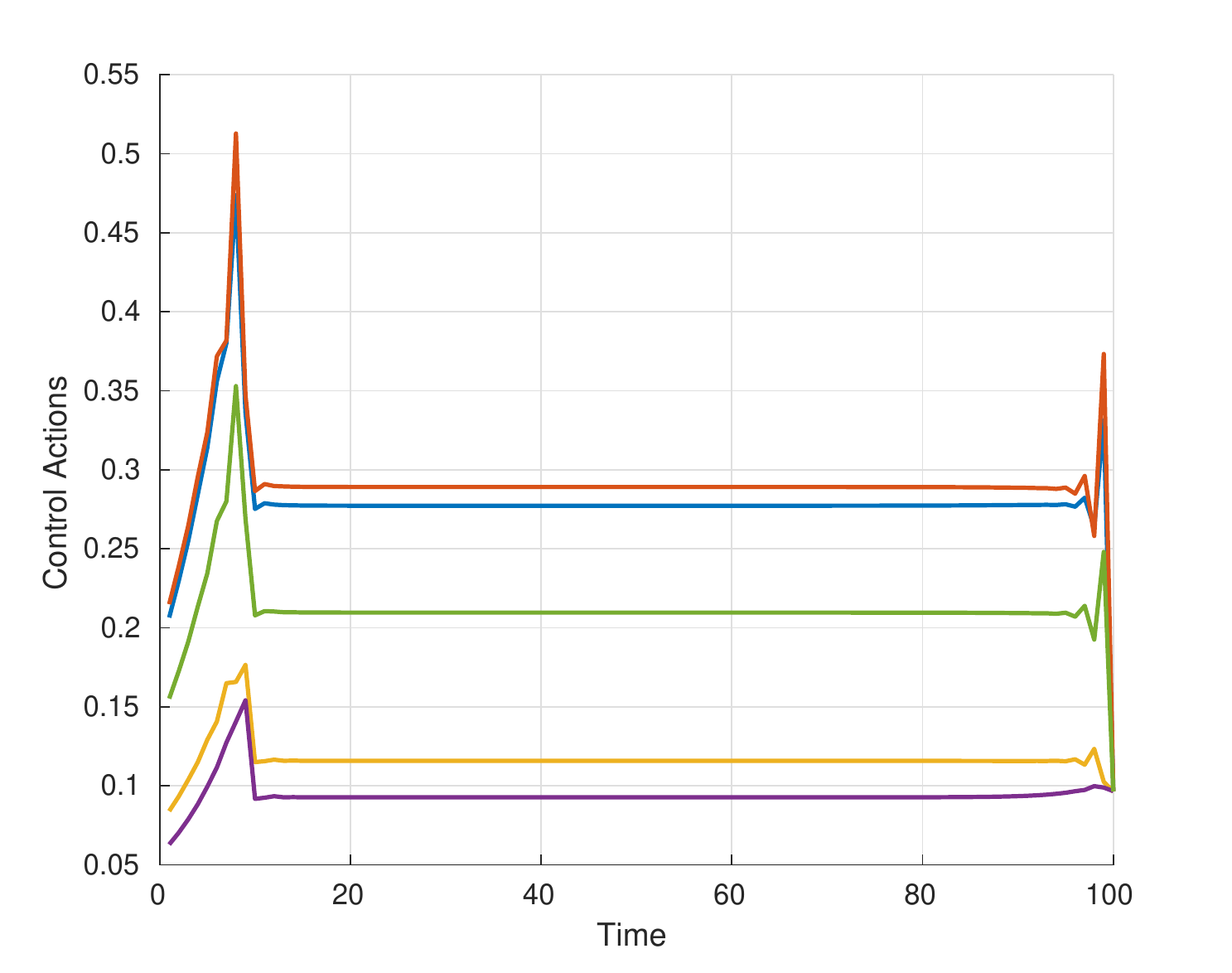}}
  \caption{\rev{A typical control action trajectory for the Group
      Ferment optimal control problem. Each curve corresponds to the
      control actions at one of $m = 5$ controlled nodes. The
      experimental setup is same as for
      Fig.~\ref{fig:grp_typical_x}.}}
  \label{fig:grp_typical_u}
\end{figure}

\section{Selecting an Optimal Set of Controlled Nodes}
\label{sec:choosing}

In the previous sections, we have studied the problem of finding the
minimum cost control when the $m$ controlled nodes are given. We now take
the design view and allow the influencing entity to choose the $m$
controlled nodes. In the discussion in this section, we restrict our
attention to the case of $R = I_m.$ Extensions to the case of a
diagonal, positive $R$ is straightforward.

The problem of choosing $m$ controlled nodes can also be seen as
designing the matrix $B$ such that $m$ of its entries are $1$ and
others are $0.$ For the optimal control problem~\eqref{eq:TF-OCP},
this reduces \rev{to} solving the following optimization problem. 

\begin{align} \label{ccn1}
  \begin{split}
    &\underset{B}{\text{minimize}} ~~ J^*_{TF}(\bar{x}) =
    \sum_{t=0}^{T-1} \|u^*(t)\|_2^2 \\
    &\text{subject to}~~ \| B \|_0 = m ,
  \end{split}
\end{align}
where, given a particular choice of $B$, $(u^*(t))_{t=0}^{T-1}$
denotes an optimal control action trajectory
for~\eqref{eq:TF-OCP}. The optimization for~\eqref{eq:GF-OCP} is
analogous with $J^*_{TF}(\bar{x})$ being replaced by
$J^*_{GF}(\bar{x}).$

\rev{It must be noted that for some graph topologies, there may exist
  sets of controlled nodes for which an equilibrium $(x^e,u^e)$ does
  not exist. That is, there are valid matrices $B^{\dagger}$ such that
  there is no solution to $x^e = Ax^e + B^{\dagger}u^e + (I_n-A)q$ for
  $x^e \geq \tau$. Towards this, we invoke Lemma~\ref{lemma:lem3} which states that an equilibrium is guaranteed to exist under Assumption~\ref{assumption:path_from_controllable}.}

%
The state-action trajectories, before the equilibrium is reached, are
difficult to analyze and this makes solving the above optimization
problem hard. Recall that the turnpike property of the optimal
solutions of~\eqref{eq:TF-OCP} and~\eqref{eq:GF-OCP} implies that the
state-action trajectory $(x^*(t),u^*(t))_{t=0}^T$ remains close to an
equilibrium point $(x^e,u^e)$, for all but a small fraction of the
time steps in $(T_0, T).$ The equilibrium point for~\eqref{eq:TF-OCP}
can be found explicitly via~\eqref{eq:equilibrium}.  Now observe that
the equilibrium $(x^e,u^e)$ is independent of the initial state
$\bar{x}$ and the final time $T$. Hence, rather than solve
\eqref{ccn1} exactly, we can consider approximately solving the
optimization problem in \eqref{ccn1}, where we optimize the choice of
controlled nodes to minimize the cost incurred while the state is
close to $x^e.$ Informally, this ensures that the cost is minimized to
keep the system running in the equilibrium state. We expect this to be
a good approximation, especially as the time horizon $T$ becomes
large. Thus the following optimization problem is a good approximation
for the exact solution of the problem of optimally selecting
controlled nodes for~\eqref{eq:TF-OCP}.
\begin{align} \label{noncvx}
  \begin{split}
    &\underset{B}{\text{minimize}} ~~  \| u^e \|_2^2 \\
    &\text{subject to} 
    \begin{cases}
      x^e = Ax^e + Bu^e + (I_n-A)q \\
      x^e \geq \tau \\
      \| B \|_0 = m.
   \end{cases}
 \end{split}
\end{align}
For problem~\eqref{eq:GF-OCP}, the second constraint is replaced by
$\phi(x^e) \geq kn.$

The sparsity constraint on $B$ in~\eqref{noncvx} makes the problem
non-convex and thus, hard to solve. The following two heuristics are
based on the above optimization problem.
\subsection{Convex Relaxation}
\label{sec:convrel}
Here we remove the constraint on $B$ in~\eqref{noncvx} and add a term
in the objective function corresponding to the $l_1$ norm of the
control, corresponding to the well-known LASSO regularization
technique \cite{tibshirani1996regression}. We define an auxiliary
variable $\tilde{u}^e \in \mathbb{R}^n$ to be used in place of $Bu^e$
such that the resulting problem formulation is:
\begin{align} \label{convex_relaxation}
 \begin{split}
   &\underset{\tilde{u}^e}{\text{minimize}}~~ \| \tilde{u}^e \|_2^2 +
   \mu \| \tilde{u}^e \|_1 \\
    &\text{subject to} 
    \begin{cases}
      x^e = Ax^e + \tilde{u}^e + (I_n-A)q \\
      x^e \geq \tau.
   \end{cases}
 \end{split}
\end{align}
Let the solution of this convex problem be $\tilde{u}^*$. The
controlled nodes are chosen as the nodes corresponding to which
$\tilde{u}^*$ has non-zero entries. The regularization coefficient
$\mu$ can be adapted to change the sparsity level desired from the
solution. However, it might not always be possible to match the
sparsity of the solution to $m$, the number of controlled nodes. In
such a scenario, one can select the top $m$ elements of $\tilde{u}^*$
to select the controlled nodes.

It must be noted here that such a convex relaxation heuristic can be
used only for problem~\eqref{eq:TF-OCP} and not for
problem~\eqref{eq:GF-OCP}. This is because the set specifying the
state constraints is convex for problem~\eqref{eq:TF-OCP} and is
generally non-convex for problem~\eqref{eq:GF-OCP}. In the following
subsection we give an heuristic which is applicable for both the
problems,~\eqref{eq:TF-OCP} and~\eqref{eq:GF-OCP}.

\subsection{A Greedy Heuristic}
\label{sec:greedy}

A greedy heuristic can be used for the non-convex
problem~\eqref{noncvx} and the corresponding
problem for~\eqref{eq:GF-OCP}. In this greedy heuristic we choose one node
per step of the algorithm. At each step we choose the node that causes
the largest reduction in the total cost of maintaining the state at
equilibrium to include into the set of controlled
nodes. Algorithm~\ref{algo:1} describes this heuristic formally for
problem~\eqref{eq:TF-OCP}. The algorithm for problem~\eqref{eq:GF-OCP}
will be similar except for a minor change in lines 8--10 that are
shown in Algorithm~\ref{algo:2}. 
\revv{To ensure that the intermediate steps of the algorithms have feasible solutions, we start building the set of controlled nodes with a set of nodes having direct paths to all other nodes. The existence of such a set of nodes is ensured by Assumption~\ref{assumption:path_from_controllable}.} We denote this small set of nodes by $C_{\text{init}}.$


\begin{algorithm}[htbp]\label{greedy_heuristic}
\SetAlgoLined\DontPrintSemicolon
\SetKwFunction{proc}{$equilibrium\_cost$}
\KwIn{$(A, q, \tau, m, n)$}
\KwOut{Set of controlled nodes}
 \emph{Let C be the set of controlled nodes}\\
 \DontPrintSemicolon  $C = C_{\text{init}}$	\Comment*[r]{initialization}
 \For{$i\leftarrow 1$ \KwTo $m$}{
  $n_i = \underset{j\in \mathcal{N}\backslash C}{\arg\min}~ equilibrium\_cost(C \cup \{j\})$\\
  \DontPrintSemicolon $ C \leftarrow C \cup n_i $  \Comment*[r]{greedy update}
}
  \setcounter{AlgoLine}{0}
  \SetKwProg{myproc}{Procedure}{}{}
  \myproc{\proc{S}}{
  $B = 0_{m \times n}$\\ 
  \DontPrintSemicolon  $i = 1$ \Comment*[r]{counter variable}
   \For{$j \in S$}
   {
 \DontPrintSemicolon  $B(j,i) = 1$\Comment*[r]{designing $B$ using $S$}
  $i = i+1$ \\
   }
   
  $  \text{minimize} ~~ \| u^e \|_2^2 $\\
  $ \text{subject to} $
    $\begin{cases}
       x^e = Ax^e + Bu^e + (I_n-A)q \\
       x^e \geq \tau
    \end{cases}$\\
\KwRet{$ \| u^e \|_2^2 $}}

  \caption{Greedy heuristic for controlled node selection for
    problem~\eqref{eq:TF-OCP}} 
 \label{algo:1}
\end{algorithm} 

\begin{algorithm}[htbp]
  \label{algo:2}
$    \underset{u}{\text{minimize}}~~ J_{GF}(\bar{x},u) = \sum_{t=0}^{T-1}  \|u(t)\|^2_2 $\\ 
$    \text{subject~to}$
$    \begin{cases}      
      \text{state~dynamics~\eqref{eq:dynamics}},\\ 
      \text{state~constraints~\eqref{eq:group-ferment}} \\ 
      x(0) = \bar{x} \in \mathbb{R}^n \text{~(given).}
    \end{cases}$\\
  \KwRet{$ J_{GF}(\bar{x},u) $}
  
  \caption{Lines 8--10 in Algorithm~\ref{algo:1} should be replaced by
    this segment to obtain the $m$ controlled nodes for
    problem~\eqref{eq:GF-OCP} using the greedy heuristic.}
  
\end{algorithm}

\section{Max Min Optimal Control}
\label{sec:maxmin}

To solve the problem~\eqref{eq:MF-OCP}, we need to define the
following two system variables in addition to $x(t) \in \mathbb{R}^n,$
the opinion at each of the nodes at time $t.$

\begin{itemize}
\item $r(t)\in \mathbb{R}$ is the \emph{record} of $\psi(x(\cdot))$ up
  to time $t,$ i.e., it is the minimum value of $\psi(x(\cdot))$ till
  time $t-1.$ Formally,
  \begin{align}
    r(t) = \begin{cases}
      \psi(x(0))  & t= 0 \\
      \min_{0 \leq \tau \leq t-1}~~\psi(x(\tau)) & t = 1, \ldots, T,
    \end{cases}
  \end{align}
\item $y(t) \in \mathbb{R},$ is the running sum of the control cost
  till time $t-1,$ i.e.,
  \begin{align}
    y(t) = \begin{cases}
      0 & t= 0 \\
      \sum_{\tau=0}^{t-1} c(u(\tau))& t = 1, \ldots, T,
    \end{cases}  
  \end{align}
  subject to the budget constraint, imposed by the boundary
  condition  
  \begin{align}
    y(T) \le C.
  \end{align}
\end{itemize}

We can see that problem~\eqref{eq:MF-OCP} is equivalent to maximizing
$r(T)$ subject to the budget constraint $y(T) \le C$ and the given
initial conditions. Thus we can use $x(t),$ $r(t),$ and $y(t)$ as the
state variables in an optimal control problem. The dynamics of these
state variables will be as follows.
\begin{align}
  \begin{split}
    x(t+1) &= f(x(t), u(t)) ~~~~~~~~~~ \text{for~} t = 0,\ldots, T-1, \\
    r(t+1) &= \min\{r(t),\psi(x(t))\} ~~ \text{for~} t = 0,\ldots, T-1, \\
    y(t+1) &= y(t) + c(u(t)) ~~~~~~~~ \text{for~} t = 0,\ldots, T-1.
  \end{split}
  \label{eq:maxmin-state-dynamics}
\end{align}

Observe that the dynamics of the state variable $r(t)$ are generally
not smooth. This implies that the classical PMP cannot be used to
solve the problem~\eqref{eq:MF-OCP}.

Optimal control with minimax cost has been studied before; see
\cite{Boltyanski12}, the authoritative monograph and the references
therein for a general introduction to the problem and the
issues. Here, we use some recent results on non-smooth PMP obtained in
\cite{kotpalliwar2019discrete}, and develop a numerical technique to
solve the optimal control problem~\eqref{eq:MF-OCP} exactly. The
non-smooth PMP solution based on \cite{kotpalliwar2019discrete}, is
discussed in the following subsection.

\subsection{Non-smooth PMP formulation}
Consider the problem~\eqref{eq:MF-OCP}. Let
$(x^*(t),r^*(t),y^*(t))_{t=0}^{T}$ and $(u^*(t))_{t=0}^{T-1}$ be an
optimal state-action trajectory that solves~\eqref{eq:MF-OCP}. Define
the Hamiltonian
\begin{align}
  \nonumber \mathbb{R}^n \times \mathbb{R} \times \mathbb{R} \times
  \mathbb{R}^n \times \mathbb{R} \times \mathbb{R} \times \mathbb{R}^m
  \ni (\zeta_x, \zeta_r, \zeta_y, \xi_x, \xi_r, \xi_y, \mu)\\ 
  \nonumber \mapsto H(\zeta_x, \zeta_r, \zeta_y, \xi_x, \xi_r, \xi_y,
  \mu):= \zeta_x^\intercal \left( A\xi_x + B\mu+(I_n-A)q \right)\\  
  \nonumber + \zeta_r \min\{\xi_r,\psi(\xi_x)\} + \zeta_y (\xi_y +
  c(\mu)).
\end{align}
Let the adjoint variables be given by
$(\lambda_x(t),\lambda_r(t),\lambda_y(t))_{t=0}^T$. For ease of
notation, we use $\lambda(t)$ in place of
$(\lambda_x(t),\lambda_r(t),\lambda_y(t))$ and $\lambda_{xr}(t)$ to
denote the tuple $(\lambda_x(t),\lambda_r(t))$. Similarly we use
$xr^*(t)$ to denote the tuple $(x^*(t),r^*(t))$. The non-smooth PMP
asserts that there exists a sequence $(\lambda(t))_{t=0}^T$ that
satisfy the following. 
\begin{itemize}
\item The adjoint variable $\lambda^*(t)$ does not vanish at any time.
\item The state dynamics follow 
  \begin{align}
    \begin{split}
      x^*(t+1) = \frac{\partial}{\partial \zeta_x}
      H(\lambda^*(t),x^*(t),r^*(t),y^*(t),u^*(t))\\
      r^*(t+1) = \frac{\partial}{\partial \zeta_r}
      H(\lambda^*(t),x^*(t),r^*(t),y^*(t),u^*(t))\\
      y^*(t+1) = \frac{\partial}{\partial \zeta_y}
      H(\lambda^*(t),x^*(t),r^*(t),y^*(t),u^*(t))\\
      \text{for~} t = 0,\ldots,T-1.
    \end{split}    
  \end{align}
\item The adjoint dynamics~\rev{\cite{kotpalliwar2019discrete}} follow 
  \begin{align}
    \begin{split}
      &\lambda_y(t-1) = \frac{\partial}{\partial \xi_y}
      H(\lambda^*(t),x^*(t),r^*(t),y^*(t),u^*(t)) \\
      &\left<\lambda_{xr}(t-1),v\right> \geq \mathcal{D}_{v}
      H(\lambda^*(t),\cdot,\cdot,y^*(t),u^*(t))(xr^*(t)) \\      
      &\text{for all~} v \in \mathbb{R}^{n+1},\text{~for~} t =
      1,\ldots,T-1. 
    \end{split}
  \end{align}
 \revv{ \item Since the objective is to maximize $r(T)$, the following terminal conditions are satisfied:
   \begin{align}
    \lambda_r(T) = 1, \label{eq:terminal-lambda} \\
    \lambda_x(T) = 0.
  \end{align}}
\item The following complementary slackness condition is satisfied.
  \begin{align}
    \lambda_y(T)(y^*(T)-C) = 0. \label{eq:complementary-slackness}
  \end{align}
\item The following Hamiltonian maximization condition is satisfied. 
  \begin{align}
    \frac{\partial}{\partial \mu}
    &H(\lambda^*(t),x^*(t),r^*(t),y^*(t),u^*(t)) = 0,\\
    &\rev{\text{i.e.,~}u^*(t) = \frac{1}{2\lambda^*_y(T)} R^{-1}
    B^\intercal\lambda_x^*(t).} \label{eq:u}
  \end{align}
\end{itemize}
\rev{The case of $\lambda^*_y(T) = 0$ corresponds to the existence of
  abnormal control. It must also be noted that the
  non-smooth PMP, similar to PMP, is based on first order conditions
  and does not guarantee the uniqueness of the solution obtained.} It
must be noted here that the non-smooth PMP uses directional
derivatives to specify the dynamics of the adjoint variables; see
definition below.

\begin{definition}
  (\hspace{1sp}\cite{clarke2013functional}, Section 1.4 on page 20)
  Let $g: \mathbb{R}^d \times \mathbb{R}^m \to \mathbb{R}^n$ be a
  continuous map.  For $y \in \mathbb{R}^m$ and a vector $v \in
  \mathbb{R}^d$, we denote by $\mathcal{D}_v g(\cdot, y)(x)$ the
  directional derivative of $g(\cdot,y)$ along $v$ at $x$, whenever
  the following limit exists:
  \begin{align}
    \mathcal{D}_v g(\cdot, y)(x) := \lim_{\theta \downarrow 0}
    \frac{g(x+\theta v, y) - g(x,y)}{\theta}
  \end{align}
\end{definition} 

Note that the directional derivative above is defined as a right-hand
(one-sided) limit.  If $g$ is continuously differentiable, then
$\mathcal{D}_v g(\cdot, y)(x) = \frac{\partial}{\partial x} g(x,y).v$.

In the following we develop a modified forward-backward sweep
algorithm to find a numerical solution of problem~\eqref{eq:MF-OCP}
satisfying the conditions of the non-smooth PMP. In the next
subsection, we outline the algorithm.

\subsection{Numerical Algorithm}
Before we discuss the algorithm, we analyze the directional derivative
of the Hamiltonian for problem~\eqref{eq:MF-OCP}. We denote the ordered
pair of vector $v_x \in \mathbb{R}^n$ and scalar $v_r \in \mathbb{R}$
by $v,$ such that $ \mathbb{R}^n \times \mathbb{R} \ni v = (v_x,
v_r)$.
\begin{align}
  \mathcal{D}_v H(\lambda,\cdot,\cdot,y,u)(xr) \nonumber\\
  = \lim_{\theta \downarrow 0}\frac{ H(x+\theta v_x, r+\theta v_r, y,
    u) - H(x, r, y, u)}{\theta} \label{eq:ddH}
\end{align}
Here we consider three cases for the the relation between $r(t)$ and
$\psi(x(t))$ to determine the directional derivatives which will then
give the adjoint dynamics.
\begin{itemize}
\item \textbf{Case 1: } for some time instant $t = t_1,$ we have
  $r(t_1) < \psi(x(t_1))$. In this case, the record state $r(t_1+1)$
  is updated such that $r(t_1+1) = r(t_1)$. The non-smooth PMP
  formulation is same as the usual PMP because the Hamiltonian is
  differentiable in all the state variables. The adjoint dynamics at
  this time instant are given by
  \begin{align*}
    \lambda_y(t_1-1) &= \lambda_y(t_1), \\
    \lambda_r(t_1-1) &= \lambda_r(t_1), \\
    \lambda_x(t_1-1) &= A^\intercal\lambda_x(t_1).
  \end{align*}
\item \textbf{Case 2: } For some time instant $t = t_2,$ we have
  $r(t_2) > \psi(x(t_2))$. In this case, the record state $r(t_2+1)$
  is updated such that $r(t_2+1) = \psi(x(t_2))$. The non-smooth PMP
  formulation is same as the usual PMP because the Hamiltonian is
  differentiable in all the state variables. The adjoint dynamics at
  this time instant are given by
  \begin{align*}
    \lambda_y(t_2-1) &= \lambda_y(t_2), \\
    \lambda_r(t_2-1) &= 0, \\    
    \lambda_x(t_2-1) &= A^\intercal\lambda_x(t_2) + \lambda_r(t_2)
    \frac{d}{dx} \psi(x(t_2)).
  \end{align*}
\item \textbf{Case 3: } For some time instant $t = t_3,$ we have
  $r(t_3) = \psi(x(t_3))$. In this case the non-smooth PMP formulation
  is not the same as the usual PMP because the Hamiltonian is not
  differentiable in all the state variables. Using the directional
  derivatives, the adjoint dynamics at the time instant $t_3$ are
  given by
  \begin{align*}
    &\lambda_y(t_3-1) = \lambda_y(t_3), \\
    &\left<\lambda_r(t_3-1),v_r\right> + \left<\lambda_x(t_3-1),v_x
    \right> \\    
    &\geq \mathcal{D}_v
    H(\lambda(t_3),\cdot,\cdot,y(t_3),u(t_3))(xr(t_3)) \\    
    &\text{for all~} v_x \in \mathbb{R}^{n} \text{~and all~} v_r \in
    \mathbb{R}.
  \end{align*}
  The directional derivative, in direction $v$, at the time $t_3$
  instant can be given by:
  \begin{align}
    \lambda_x^\intercal(t_3) A v_x + \lambda_r(t_3) \min\left\{ \left<
    \frac{d}{dx}\psi(x(t_3)),v_x \right>, v_r \right\}.\label{eq:dd}
  \end{align}
 \rev{ A proof of~\eqref{eq:dd} is in Appendix~\ref{appendix:2}. }
  
  The adjoint dynamics are:
  \begin{align}
    &\lambda_y(t_3-1) = \lambda_y(t_3), \nonumber\\
    &\left<\lambda_r(t_3-1),v_r\right> + \left<\lambda_x(t_3-1),v_x
    \right> \nonumber\\    
    &\geq \lambda_x^\intercal(t_3) A v_x + \lambda_r(t_3) \min\left\{
    \left< \frac{d}{dx}\psi(x(t_3)),v_x \right>, v_r
    \right\} \label{eq:directional-derivatives} \\
    &\text{for all~} v_x \in \mathbb{R}^{n} \text{~and all~} v_r \in
    \mathbb{R}. \nonumber
  \end{align}
  Here we consider the following two possibilities:
  \begin{align*}
    \left< \frac{d}{dx}\psi(x(t_3)),v_x \right> \geq v_r,\\
    \left< \frac{d}{dx}\psi(x(t_3)),v_x \right> < v_r.
  \end{align*}

  We need $\lambda_x(t_3-1)$ and $\lambda_r(t_3-1)$ to be such
  that~\eqref{eq:directional-derivatives} is satisfied for both the
  above stated possibilities. We use the following values of
  $\lambda_x(t_3-1)$ and $\lambda_r(t_3-1)$:
  \begin{align}
    \begin{split}
      \lambda_x(t_3-1) &= A^\intercal \lambda_x(t_3) +\phi\lambda_r(t_3)
      \frac{d}{dx}\psi(x(t_3)),\\      
      \lambda_r(t_3-1) &=  (1-\phi)\lambda_r(t_3), \\
      \text{for some~} \phi &\in[0,1].
    \end{split} \label{eq:adjoint-dynamics}
  \end{align}

  We now verify that this indeed
  satisfies~\eqref{eq:directional-derivatives}.

  When $\left< \frac{d}{dx}\psi(x(t_3)),v_x \right> >
  v_r,$~\eqref{eq:directional-derivatives} reduces to:
  \begin{align}
    &\left<\lambda_r(t_3-1),v_r\right> + \left<
    \lambda_x(t_3-1),v_x \right> \nonumber\\
    &\geq \lambda_x^\intercal(t_3) A v_x + \lambda_r(t_3)  v_r. \label{case1} 
  \end{align}
  Using the adjoint dynamics given by~\eqref{eq:adjoint-dynamics}, we
  see that
  \begin{align}
    &\left<\lambda_r(t_3-1),v_r\right> + \left< \lambda_x(t_3-1),v_x
    \right> = \left<(1-\phi)\lambda_r(t_3),v_r\right> \nonumber\\
    & + \left< A^\intercal \lambda_x(t_3) + \phi\lambda_r(t_3)
    \frac{d}{dx}\psi(x(t_3)),v_x \right> \nonumber \\    
    & = \lambda_x^\intercal(t_3) A v_x + (1-\phi) \lambda_r(t_3) v_r  \nonumber\\
   & +  \phi\lambda_r(t_3) \left< \frac{d}{dx}\psi(x(t_3)),v_x \right> \nonumber \\    
    & \geq \lambda_x^\intercal(t_3) A v_x + \lambda_r(t_3) v_r
    \nonumber \\    
    &\text{for~} \phi\lambda_r(t_3) \geq 0,
  \end{align}
  thus satisfying~\eqref{case1} if $\phi\lambda_r(t_3) \geq 0$.

  Now we consider the second case, i.e., when $\left<
  \frac{d}{dx}\psi(x(t_3)),v_x \right> < v_r$. In this case,
  condition \eqref{eq:directional-derivatives} reduces to:
  \begin{align}
    &\left<\lambda_r(t_3-1),v_r\right> + \left<
    \lambda_x(t_3-1),v_x \right> \nonumber\\
    &\geq \lambda_x^\intercal(t_3) A v_x + \lambda_r(t_3) \left<
    \frac{d}{dx}\psi(x(t_3)),v_x \right> . \label{case2}
  \end{align}  

  From the adjoint dynamics in~\eqref{eq:adjoint-dynamics}, we see that
  \begin{align}
    &\left<\lambda_r(t_3-1),v_r\right> + \left< \lambda_x(t_3-1),v_x
    \right> = \left<(1-\phi)\lambda_r(t_3),v_r\right> \nonumber\\
    & + \left< A^\intercal \lambda_x(t_3) + \phi\lambda_r(t_3)
    \frac{d}{dx}\psi(x(t_3)),v_x \right> \nonumber \\  
    & = \lambda_x^\intercal(t_3) A v_x + (1-\phi) \lambda_r(t_3) v_r  \nonumber \\& + \phi\lambda_r(t_3) \left< \frac{d}{dx}\psi(x(t_3)),v_x \right>   \nonumber \\  
    & \geq \lambda_x^\intercal(t_3) A v_x + \lambda_r(t_3) \left<
    \frac{d}{dx}\psi(x(t_3)),v_x \right> \nonumber \\  
    &\text{for~}  (1-\phi)\lambda_r(t_3) \geq 0, 
  \end{align}
  
  thus satisfying~\eqref{case2} if $(1-\phi)\lambda_r(t_3) \geq 0$. \revv{To satisfy both~\eqref{case1}
  and~\eqref{case2}, we need to satisfy both $\phi\lambda_r(t_3) \geq 0$ and $(1-\phi)\lambda_r(t_3) \geq 0$. This implies $\lambda_r(t_3) \geq \phi\lambda_r(t_3) \geq 0$ and therefore $\phi \in [0,1]$. This verifies that our design in~\ref{eq:adjoint-dynamics} satisfies~\ref{eq:directional-derivatives}. Note that this is valid only for $\lambda_r(t_3) \geq 0,$ which follows the terminal conditions~\eqref{eq:terminal-lambda} and the dynamics of $\lambda_r(\cdot)$ ensuring that it doesn't change its sign.} The exact value of $\phi$ is determined by the algorithm to ensure the required relation between $r(t_3)$ and $\psi(x(t))$.
\end{itemize}

The detailed algorithm is described in Algorithm~\ref{algo:maxmin}.
The value of the adjoint variable $\lambda_y(T)$ is determined by the
shooting method, implemented in the outer loop of
Algorithm~\ref{algo:maxmin}, such that the complementary slackness
conditions~\eqref{eq:complementary-slackness} are satisfied.

\begin{algorithm}[htbp] \label{algo:maxmin}
\SetAlgoLined\DontPrintSemicolon \SetKwFunction{proc}{$Backward$}
\KwIn{$A, B, Budget,\bar{x}, \varepsilon_1, \varepsilon_2,w,\mu$}
\KwOut{Optimal state-action trajectory $(x^*,u^*)$} \emph{Set: }
$\lambda_x(T) = 0,$ $\lambda_r(T) = 1$.\\ \emph{Initialize: }
$\lambda_y^{(0)}(T) =1.$ \\ \Repeat {$p^* \text{such that~}
  \left|\lambda_y^{(p^*)}(T) - \lambda_y^{(p^*-1)}(T)\right| <
  \varepsilon_2$} { \emph{Initialize: } $u^{(0)} = 0$. \\ \Repeat
  {$j^* \text{such that~} \left\|u^{(j^*)} - u^{(j^*-1)}\right\|_F <
    \varepsilon_1$} { \emph{Forward Sweep: } Find $x^{(j)}$ using
    $u^{(j)}$ via~\eqref{eq:maxmin-state-dynamics}.\\ \emph{Backward
      Sweep: }\\$\tilde{u} = Backward\left(
    \lambda_x(T),\lambda_r(T),\lambda_y^{(p)}(T),x^{(j)}\right)$\\ $u^{(j+1)}
    = wu^{(j)} + (1-w)\tilde{u}$\\} $Expenditure = \sum_{t=0}^{T-1}
  c\left(u^{(j^*)}(t)\right)$\\ $\lambda_y^{(p+1)}(T) =
  \lambda_y^{(p)}(T) - \mu (Expenditure - Budget)$\\ } $u^*
=u^{(j^*)},$ $x^*$ from $u^*$ via~\eqref{eq:maxmin-state-dynamics}.
\\ \SetKwProg{myproc}{Procedure}{}{}
\myproc{\proc{$\lambda_x(T),\lambda_r(T),\lambda_y^{(p)}(T),x^{(j)}$}}{
  \For{$\tau\gets T-1$ \KwTo $0$}{ Find $\phi$ such that \rev{$\psi(x(\tau)) =
    \psi(x(\tau+1))$,} \\ with adjoint
    dynamics~\eqref{eq:adjoint-dynamics} and
    control~\eqref{eq:u}.\\ \uIf{$\phi>1$} {$\lambda_x(\tau) =
      A^\intercal\lambda_x(\tau+1) + \lambda_r(\tau+1) \left<
      \frac{d}{dx}\psi(x(\tau+1)) \right> $\\ $\lambda_r(\tau) =
      0$\\ $\lambda_y(\tau) = \lambda_y(\tau+1)$} \uElseIf{$\phi<0$}
    {$\lambda_x(\tau)=A^\intercal\lambda_x(\tau+1)
      $\\ $\lambda_r(\tau) = \lambda_r(\tau+1)$\\ $\lambda_y(\tau) =
      \lambda_y(\tau+1)$} \Else
    {$\lambda_x(\tau)=A^\intercal\lambda_x(\tau+1) +
      \phi\lambda_r(\tau+1) \left< \frac{d}{dx}\psi(x(\tau+1)) \right>
      $\\ $\lambda_r(\tau) =
      (1-\phi)\lambda_r(\tau+1)$\\ $\lambda_y(\tau) = \lambda_y(\tau+1)$}
   \rev{ $\tilde{u}(\tau) = \frac{1}{2\lambda_y^{(p)}(T)} R^{-1} B^\intercal\lambda_x(\tau)$}
  } \KwRet{$\tilde{u}$}}
 \caption{Numerical algorithm to solve the maxmin ferment problem}
\end{algorithm}

\section{Numerical Results}
\label{sec:sim}

In this section we report the results from a subset of the extensive
numerical experiments that we performed. For the~\eqref{eq:TF-OCP}
problem, we first study the performance of our algorithm to choose the
controlled nodes and compare it against some natural heuristics. We
show that our algorithm is better. We then study the effect on the
cost of maintaining ferment as a function of different
parameters---structure of the network, the threshold level, the number
of controlled nodes, and the stubbornness of the nodes. For all
these experiments, the results for the~\eqref{eq:GF-OCP} problem are
qualitatively similar to that of~\eqref{eq:TF-OCP} and we can draw
similar conclusions. Hence we do not report those results. However, we
do study the effect of parameters $a$ and $k$ on the cost of
maintaining ferment and also analyze the opinion levels at the nodes
in the network. Finally, for the~\eqref{eq:MF-OCP} problem, we study
the behavior of the minimum value that is attained as a function of
the budget.

The setup for the experiments is as follows. For all the experiments,
the influence matrix $A$ is such that the sum of each row of $A$ is
$0.9$, making it sub-stochastic. Further, each agent places equal
weight\footnote{The experiments studying the effect of node
  stubbornness in \ref{sec:stubborn} do not follow this rule.} on its
opinion and that of each of its neighbors. This means that for any
node $i$ and each of its neighbors $j$, $a_{ii}=a_{ji}=0.9/(N_i+1)$
where $N_i$ is the number of in-neighbors of $ i.$ The initial
opinions are taken as $\bar{x} = 0.5$ and the quiescent level is set
to $q=0$ such that the opinion recedes to zero in the absence of
external control. The threshold level $\tau_i$ is set to $0.7$ for all
$i.$ The system is studied for $T = 100$ time steps and the threshold
level is enforced from $T_0 = 10$ onwards. For all the experiments, we
take the matrix $R$ in the cost function~\eqref{eq:cost} to be the
identity matrix $I_m.$ We use the open source software
CasADi\cite{Andersson2018} for numerical simulations. The state-action
trajectories obtained from CasADi are verified to satisfy the
necessary conditions obtained using PMP in Section~\ref{sec:PMP}. Our
experiments are performed on three kinds of random networks, each with
$n=50$ nodes---the Erd\H{o}s-R\`enyi (ER)
graphs~\cite{erdos1959random}, Barab\`asi-Albert (BA)
graphs~\cite{barabasi1999emergence}, and $k$-regular graphs ($k$R).
Finally, the results are averaged over $100$ realizations of the
random network topology. \rev{The standard deviations in the results
  are also indicated on the plots. The height of the bar on one side
  of the point is equal to the standard deviation in $100$ random
  realizations.

  It must noted that the initial conditions of the states are not
  crucial to the results of the simulation tests. As seen earlier in
  the paper, the equilibrium state-control trajectories are
  independent of the initial conditions. For $T$ large in comparison
  to $T_0$, the initial conditions have negligible effect on the total
  costs of maintaining ferment.}

\subsection{Total Ferment ~\eqref{eq:TF-OCP}}
We begin by studying the impact of several model parameters on the
cost of the~\eqref{eq:TF-OCP} problem.

\subsubsection{Choosing controlled nodes}
Social network literature provides several notions of centrality and
we investigate the performance when such `central nodes' are used as
controlled nodes. We consider two such criteria
\begin{itemize}
\item Nodes with high out-degrees are clearly more influential than
  those with lower out-degrees. \textit{Degree centers} of a graph is
  the set of nodes with the highest out-degree.
\item Nodes that are close to a large number of nodes can propagate a
  control signal quicker than others. \textit{Distance center} of a
  graph is the set of nodes that have the minimum eccentricity
  (largest hop-distance from the node to any other node).
\end{itemize}
We compare the cost of maintaining ferment when the set of $m$
controlled nodes are chosen to be the degree centers, the distance
centers, and chosen using the convex relaxation described in
Section~\ref{sec:convrel} and using the greedy algorithm described in
Section~\ref{sec:greedy}. Finding the distance center is in general
NP-hard, and we instead use a greedy approximation algorithm
\cite{hochbaum1985best}. The results for the three kinds of random
networks, each with $n=50$ nodes and an average degree of $6$, are
tabulated in Tables~\ref{tab:2}--Table~\ref{tab:4}.

\begin{table}[htbp]
  \begin{tabular}{ |p{1.2cm}||p{0.9cm}|p{0.9cm}|p{0.9cm}|p{0.9cm}|p{0.9cm}|  }
    \hline
    Method & $m=2$ & $m=4$ & $m=6$ & $m = 8$ & $m=10$\\
    \hline
    Greedy  &  174.94  & 95.90 &  71.01  & 58.13 &  48.81 \\
    Convex  &   181.38 & 105.45 & 83.67 &  72.31 &  65.08 \\ 
    Degree centers & 180.36 & 102.18 & 76.79 &  63.33  & 54.69\\
    Distance centers  & 1392.05 & 707.10 & 478.16 & 358.00 & 287.09 \\
    \hline
  \end{tabular}
  \vspace{0.05cm}
  \caption{Cost of maintaining ferment in BA networks.}
  \label{tab:2}
\end{table}

\begin{table}[htbp]
 \begin{tabular}{ |p{1.2cm}||p{0.9cm}|p{0.9cm}|p{0.9cm}|p{0.9cm}|p{0.9cm}|  }
    \hline
    Method & $m=2$ & $m=4$ & $m=6$ & $m = 8$ & $m=10$\\
    \hline
    Greedy            &607.18 &276.57 &181.24 &136.59 &110.15 \\
    Convex            &669.55 &311.03 &195.97 &142.31 &120.44 \\ 
    Degree centers    &680.96 &325.52 &216.67 &166.65 &137.98 \\
    Distance centers  &3730.61 &1606.88 &921.27 &626.96 &451.36 \\
    \hline
  \end{tabular}
  \vspace{0.05cm}
  \caption{Cost of maintaining ferment in a ER networks.}
   \label{tab:3}
\end{table}

\begin{table}[htbp]
 \begin{tabular}{ |p{1.2cm}||p{0.9cm}|p{0.9cm}|p{0.9cm}|p{0.9cm}|p{0.9cm}|  }
    \hline
    Method & $m=2$ & $m=4$ & $m=6$ & $m = 8$ & $m=10$\\
    \hline
    Greedy            &1535.32 &645.99 &397.86 &284.13 &220.00 \\
    Convex            &2003.55 &825.47 &481.33 &326.47 &251.48 \\ 
    Degree centers    &1961.74 &817.46 &502.07 &346.07 &261.73 \\
    Distance centers  &1811.33 &761.41 &471.87 &336.70 &256.54 \\
    \hline
  \end{tabular}
  \vspace{0.05cm}
  \caption{Cost of maintaining ferment in $k$R networks.}
   \label{tab:4}
\end{table}

From the Tables~\ref{tab:2}--Table~\ref{tab:4}, we see that the greedy
algorithm performs the best in all cases. It is also observed that the
distance centres performs very poorly for the BA and ER graphs, while
its performance is comparable to the other heuristics for the $k$R
graph. One reason why distance centres perform so badly for the BA and
ER graphs is that it sometimes fails to pick the nodes with the larger
out-degrees, which can have large influence in the network. In the
$k$R graphs, this heuristic is good because all nodes have equal
out-degrees. On the other hand, the degree centers heuristic picks the
nodes with the highest out-degrees and thus performs fairly
well. However, it is not the best scheme as it does not account for
the distance to other nodes while selecting the controlled
nodes. The convex relaxation heuristic also performs comparably to the
degree centers but not as good as the greedy heuristic. Given these
results, in all of the following experiments, we will use the greedy
algorithm for selecting the controlled nodes.

\subsubsection{Effect of Network Structure} 

\begin{figure}
  \centering{
    \includegraphics[width=0.8\columnwidth,height =0.6\columnwidth ]{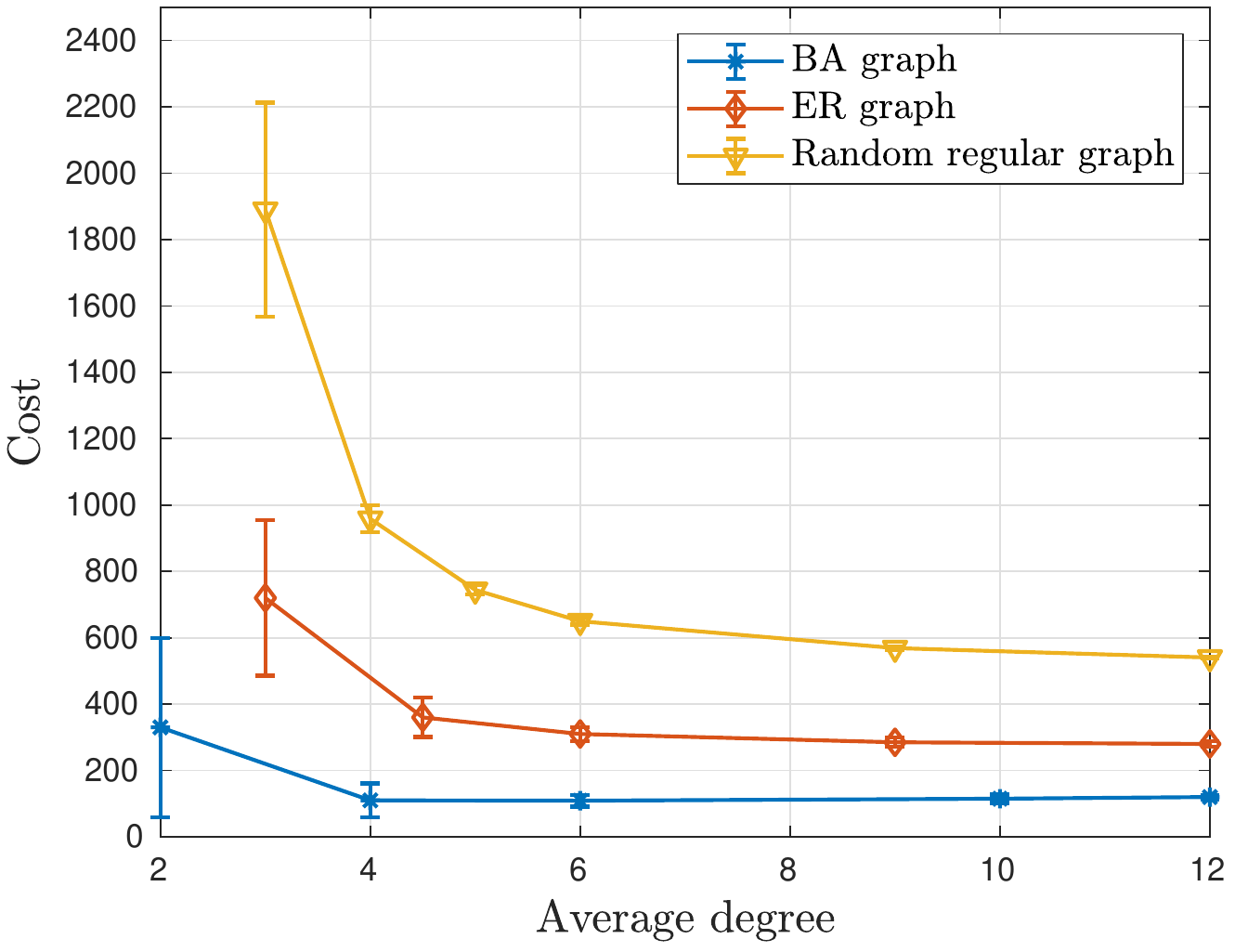}
  }
  \caption{Cost of maintaining ferment as a function of average
    degree.}
  \label{fig:deg1}
\end{figure}

Here we study the variation of the cost of maintaining ferment with
the average degree and the degree distribution of the network. The
variation in the degree distributions is provided by the different
networks. For each given average degree, we consider the BA graphs
with the power law degree distribution, the Erd\H{o}s-R\`enyi (ER)
graphs~\cite{erdos1959random} with binomial degree distribution and
the random regular graphs with constant degree. Each network consists
of $n = 50$ nodes with $m=5$ controlled nodes, chosen according to
the greedy heuristic presented in Section~\ref{sec:greedy}. We plot
the results in Fig.~\ref{fig:deg1}.

From Fig.~\ref{fig:deg1}, we see that it is cheaper to maintain
ferment in a network with a higher average degree but the marginal
gain decreases rather rapidly with increasing degree. This is to be
expected since once the graph is fairly well-connected, all nodes are
easily accessible and increasing the degree further has minimal
benefit. For a given average degree, we also see that the cost is the
least for the BA network and and highest for $k$R graphs, indicating
that the presence of nodes with high degrees makes it easier to
maintain the ferment.


\subsubsection{Effect of Threshold Value} 
\begin{figure}
  \centering
      {\includegraphics[width=0.8\columnwidth,height =0.6\columnwidth]{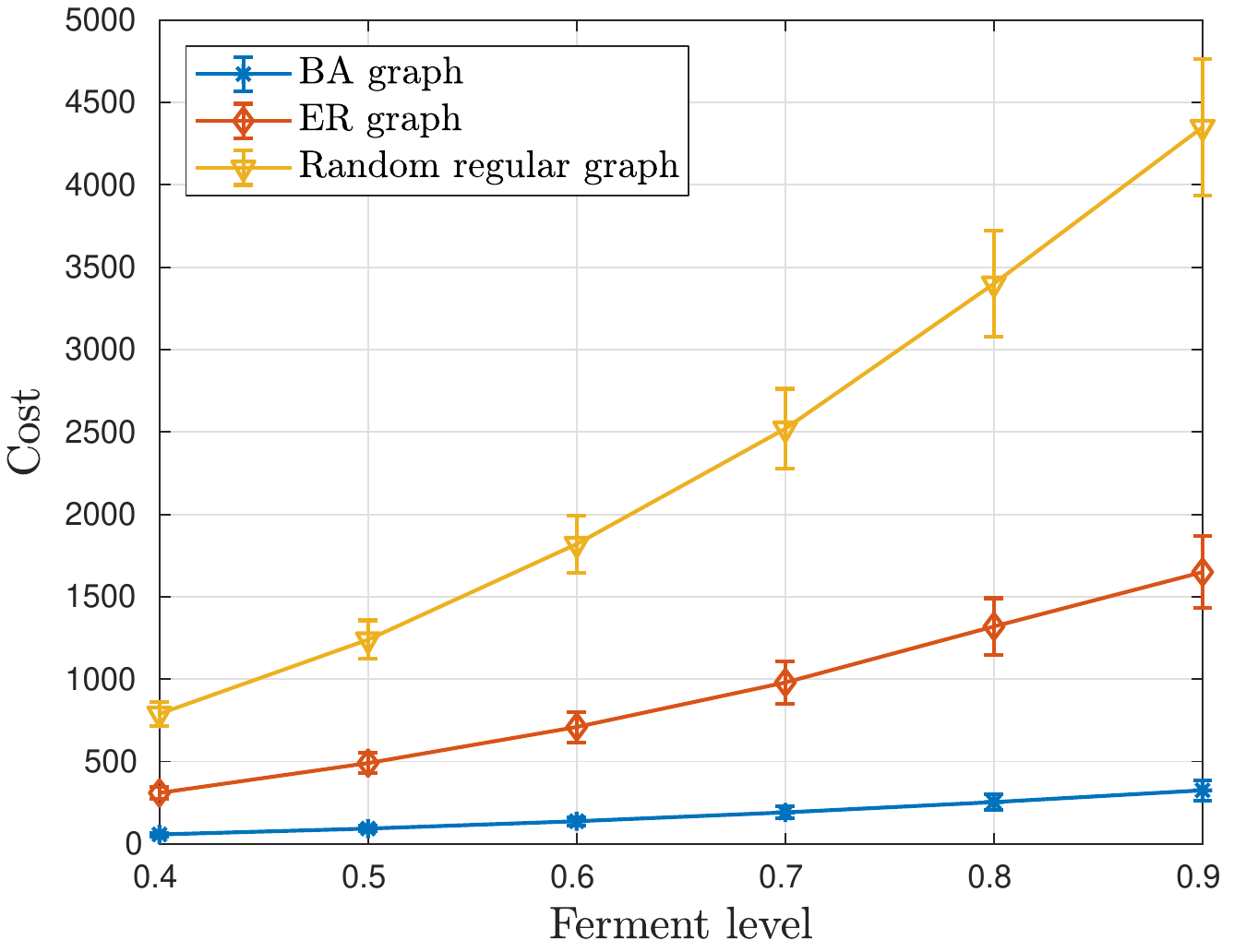}}
      \caption{Cost of maintaining ferment as a function of desired
        ferment level.}
\label{fig:ferm}
\end{figure}

In Fig.~\ref{fig:ferm}, we plot the cost of maintaining ferment as a
function of the threshold $\tau.$ The cost is approximately quadratic
with respect to $\tau.$ This is to be expected since the cost function
is a quadratic. In keeping with the finding from the previous
experiment, the cost is highest in the $k$R graphs and lowest for BA
graphs. The effect of the degree distribution is noteworthy with
significant cost reductions being obtained when there are a few nodes
with large degrees as in the BA graph.


\subsubsection{Number of Controlled Nodes} 

\begin{figure}
  \centering
      {\includegraphics[width=0.8\columnwidth,height =0.6\columnwidth ]{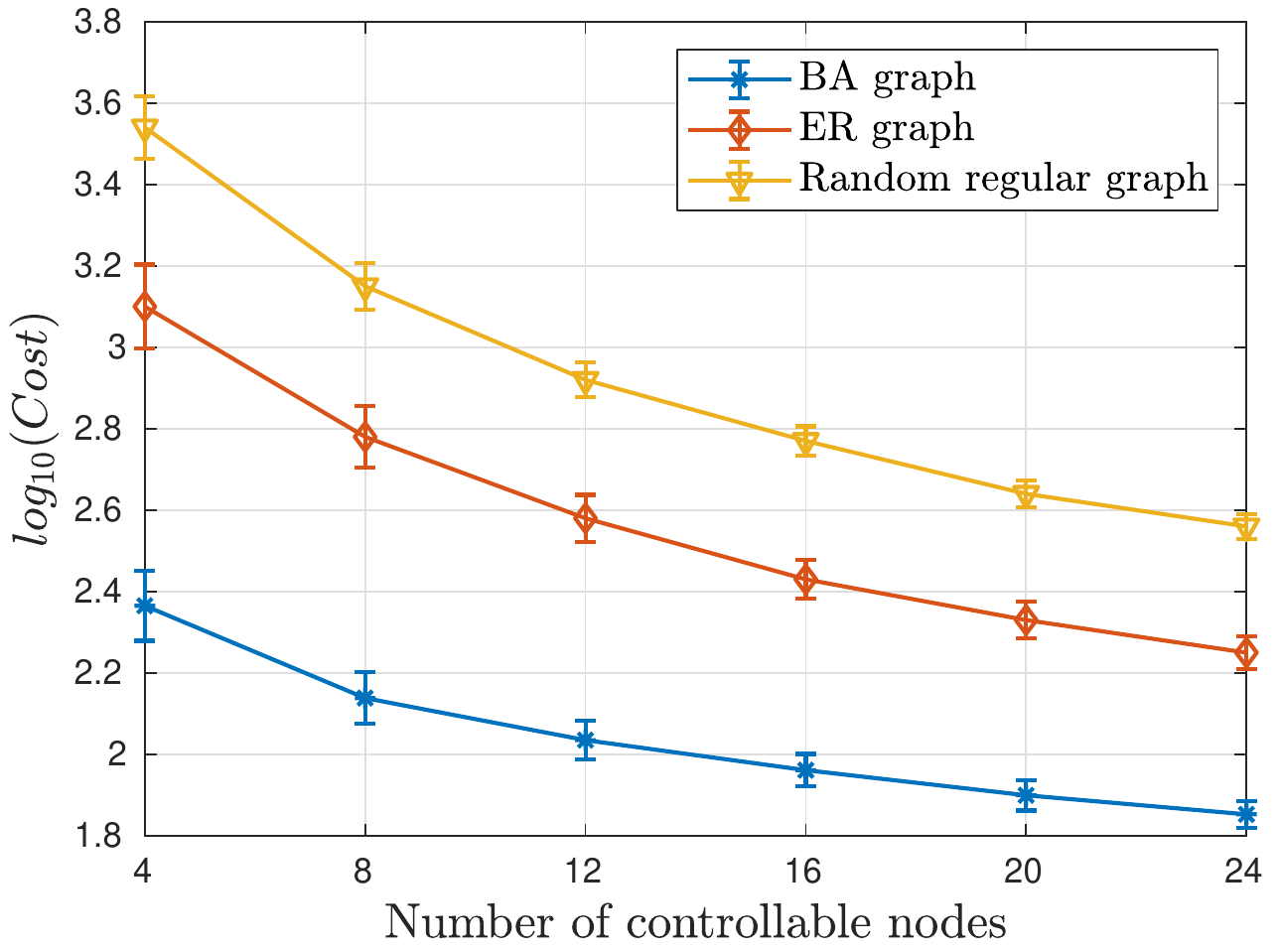}}
      \caption{Cost of maintaining ferment as a function of the number of
        controlled nodes.}
      \label{fig:cnodes}
\end{figure} 

Fig.~\ref{fig:cnodes} shows the cost of maintaining ferment as a
function of $m,$ the number of controlled nodes. Note that the cost
is plotted on a logarithmic scale indicating that the reduction in
cost is quite steep. This sharp reduction in cost is due to two
reasons: firstly, as the cost function for control inputs is convex,
having to apply smaller control inputs at a larger number of
controlled nodes decreases the total cost. The second reason is that
having a larger number of controlled nodes decreases the average
distance of nodes from a controlled node, thus helping reduce the
control inputs required to influence the most remote nodes.



\subsubsection{Effect of Stubbornness} 
\label{sec:stubborn}
\begin{figure}
  \centering
      {\includegraphics[width=0.8\columnwidth,height =0.6\columnwidth ]{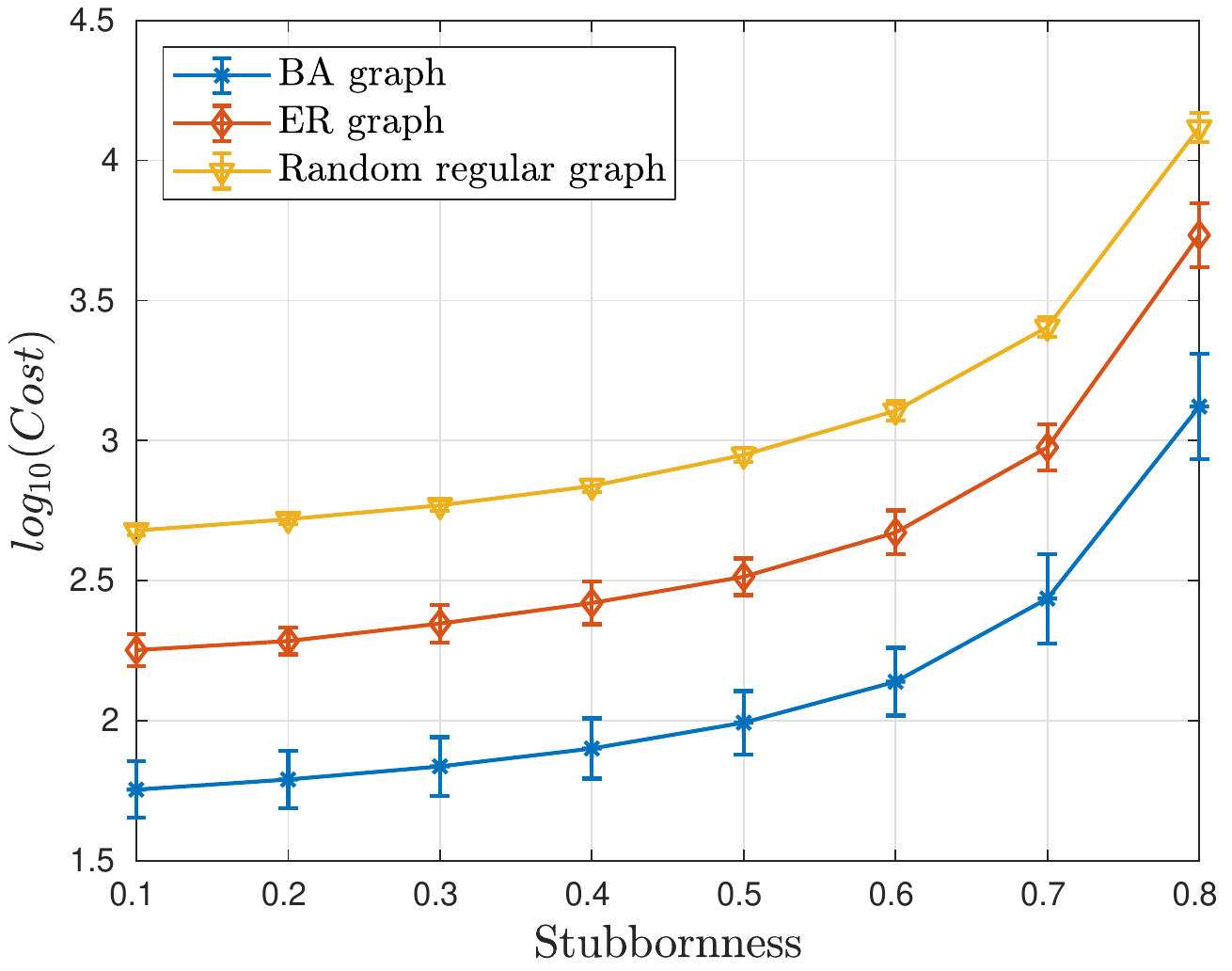}}
      \caption{Cost of maintaining ferment as a function of the
        stubbornness of agents when all the agents are stubborn.}
      \label{fig:stub}
\end{figure}

\begin{figure}
  \centering
      {\includegraphics[width=0.8\columnwidth,height =0.6\columnwidth ]{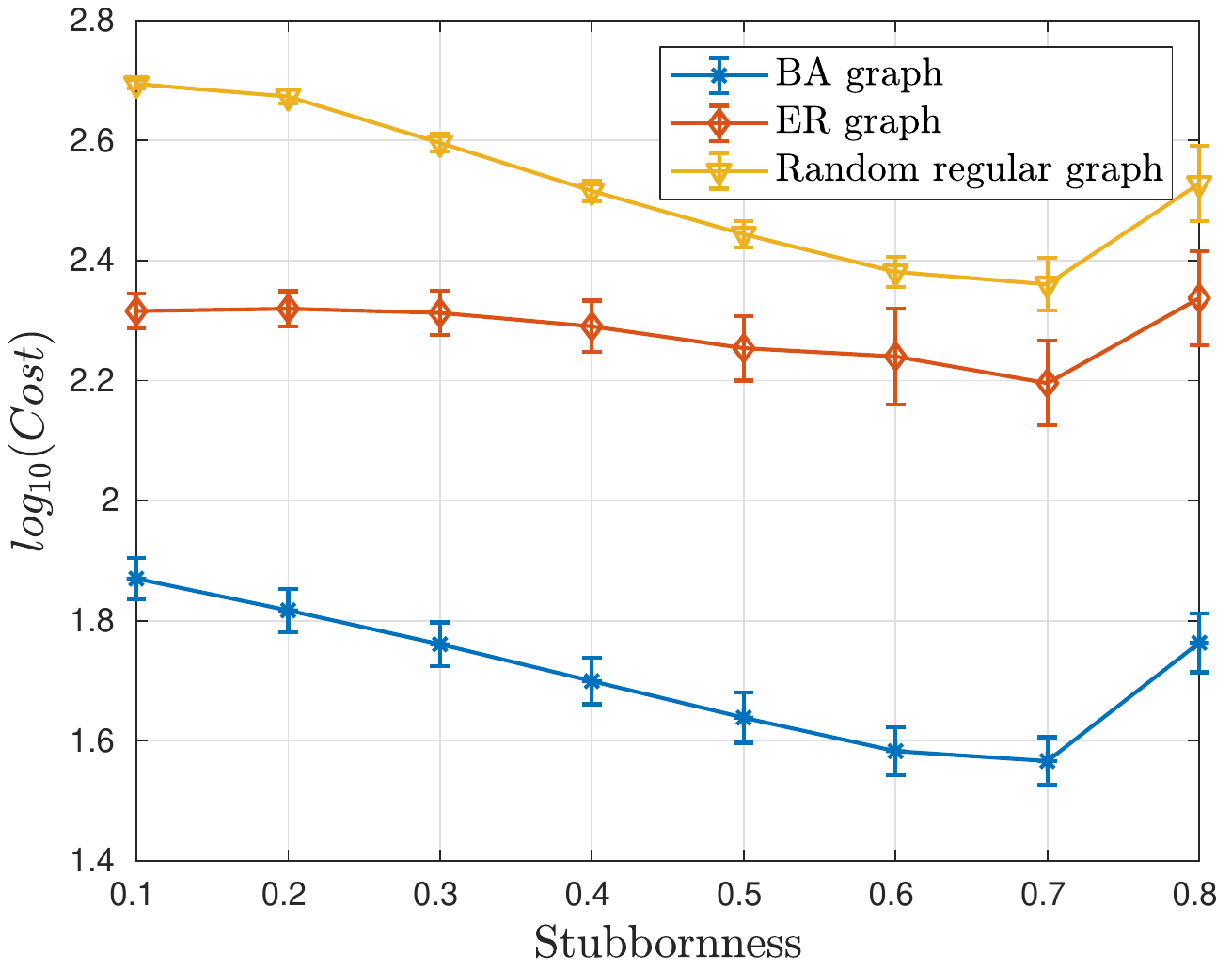}}
      \caption{Cost of maintaining total ferment: $80\%$ agents have
        $a_{ii}=0,$ plot of cost \textit{vs} $a_{ii}$ of stubborn
        nodes. }
      \label{fig:stub1}
\end{figure}

\begin{figure}
  \centering
      {\includegraphics[width=0.9\columnwidth,height =0.65\columnwidth ]{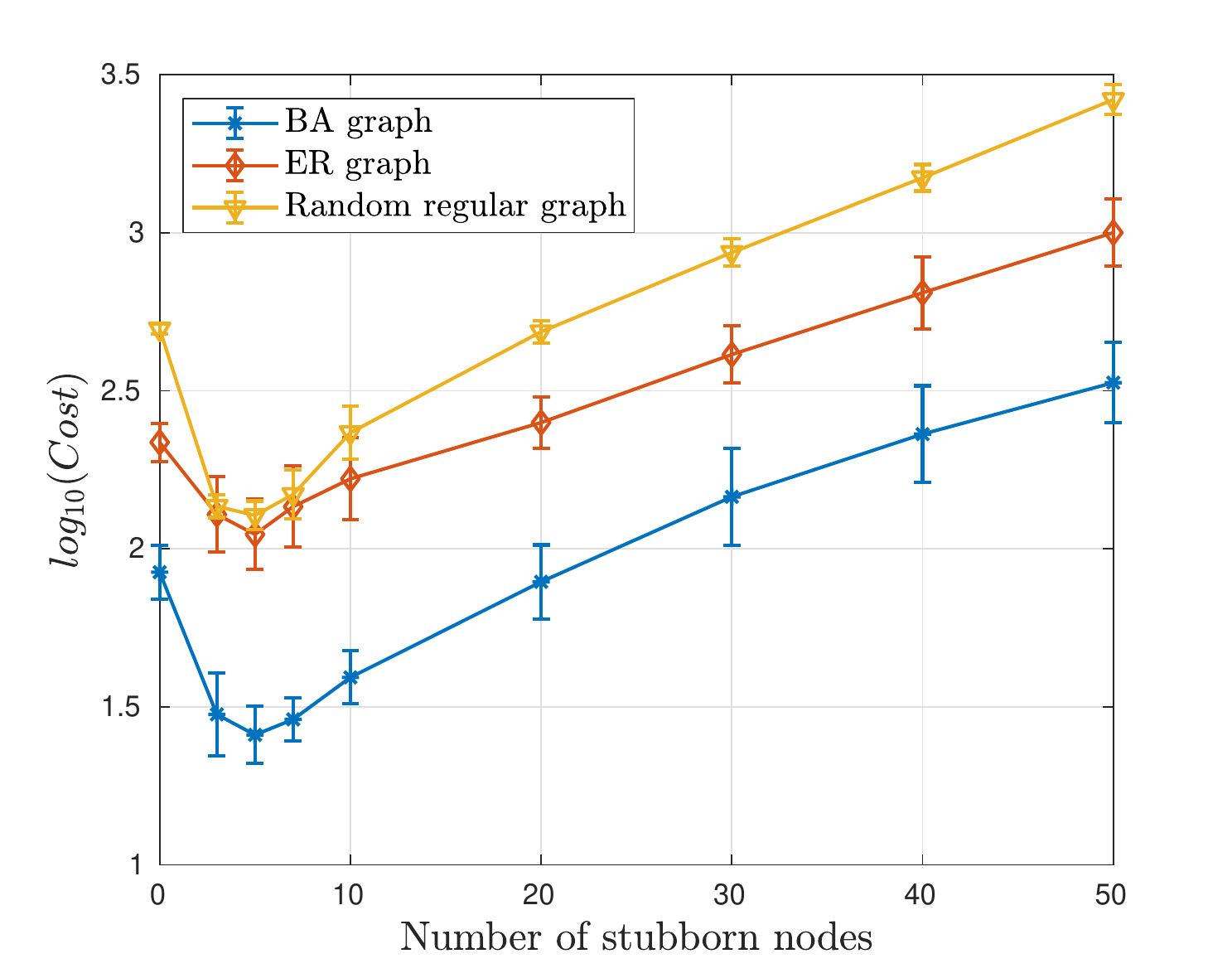}}
      \caption{Cost of maintaining ferment as a function of the number
        of stubborn agents; $a_{ii}=0.7$ for each stubborn agent.}
      \label{fig:stub2}
\end{figure}

Recall that in \eqref{eq:free-opinion-dynamics}, $a_{ii}$ denotes the
stubbornness of node $i.$ We now study its effect on the cost of
maintaining ferment.  In the first set of experiments we assume that
$a_{ii}$ is the same for all $i.$ The $a_{ji}$ are calculated as
before, i.e., $a_{ji} = (0.9-a_{ii})/N_i$ where $N_i$ is the number of
in-neighbors of $i.$ Fig.~\ref{fig:stub} plots the cost as a function
of $a_{ii}.$ We see that the cost increases very rapidly with an
increase in the stubbornness of the agents in the network. This
indicates that a thorough understanding of the population stubbornness
is vital before starting a campaign for maintaining ferment on a topic
for a certain period of time.

Next, we repeat this experiment except that now only a fraction
($20\%$) of the nodes are stubborn, chosen at random from the
network. The cost of maintaining ferment as a function of $a_{ii}$ is
plotted in Fig.~\ref{fig:stub1}. Here the results are very different
from the case of all agents being stubborn. The cost first decreases
with the level of stubbornness $a_{ii}$ and then increases. The
decrease can be explained as the stubborn nodes are good candidates
for controlled nodes. The increase, seen for a very high level of
stubbornness is because of those stubborn nodes which were not chosen
as controlled nodes. Recall that in this experiment, there are $m=5$
controlled nodes and $10$ stubborn nodes.

We also perform another variation of this experiment with a varying
fraction of the agents being stubborn. The stubbornness of the
stubborn agents is set to $a_{ii} = 0.7$, and the number of stubborn
agents is varied from $0$ to $50$. The results of this experiment are
in Fig.~\ref{fig:stub2}. It can be seen that the cost of maintaining
ferment is smallest when the number of stubborn nodes is $5$ for all
three types of networks. This is equal to the number of controlled
nodes, $m = 5$ and stubborn nodes make good controlled
nodes. \rev{This is backed by the observation that when the number of
  stubborn nodes is $5$, on average, $4.19$, $4.52$, and $4.97$
  controlled nodes are stubborn in the ER, BA, and $k$R graphs
  respectively.} The cost increases with increasing number of stubborn
nodes, beyond $5$. This is because these additional stubborn nodes are
not controlled and thus they lead to a higher cost of maintaining
ferment.
  

\subsection{Group Ferment~\eqref{eq:GF-OCP}}
We will now consider the~\eqref{eq:GF-OCP} problem. For all the
experiments conducted above, the results for the~\eqref{eq:GF-OCP}
problem are qualitatively similar to that of~\eqref{eq:TF-OCP} and we
can draw similar conclusions. Hence we do not report those results. We
will instead study the effect of parameters $a$ and $k$ on the cost of
maintaining group ferment.

\subsubsection{Effect of $a$ and $k$} 
Recall that $a$ is the slope parameter of the sigmoid function and $k$
is the fraction that determines the level of group ferment to be
maintained. As before, we take the number of nodes in the network to
be $n = 50$, the average degree is set to $6$, and the threshold
parameter $\tau$ is set to $0.7$. The system is observed for $T = 100$
time steps and the setup time $T_0 = 10$. The number of controlled
nodes is $m=5$ and they are chosen by using the greedy heuristic. As
before, the results are averaged over $100$ realizations of the
network topology. In Tables~\ref{tab:3BA},~\ref{tab:3KR},
and~\ref{tab:3ER}, we tabulate the average cost of maintaining group
ferment as a function of the parameters $a:$ slope of sigmoid function
and $k:$ the level of group ferment desired. We do this experiment for
the ER, BA, and $k$R graphs. 
   
\begin{table}[htbp]
  \begin{tabular}{ |P{1.3cm}||P{1.2cm}|P{1.2cm}|P{1.2cm}|P{1.2cm}|  }
    \hline
    $a$ $\downarrow$~~~ $k$ $\rightarrow$ & $0.5$ & $0.6$ & $0.7$ & $ 0.8$\\
    \hline
    $1$  & 44.40 &	115.77 &	 234.56 & 441.10  \\
    $3$  & 44.61 &	65.64 &	93.92 &	136.27\\ 
    $5$  & 44.97 &	57.91 &	74.24 &	97.24\\
    $7$  & 45.38 &	55.10 & 66.95 &	83.05  \\
    $10$ & 45.95 &	53.37 &	62.09 & 77.39 \\
    \hline
  \end{tabular}
  \vspace{0.05cm}
  \caption{Cost of maintaining group ferment in BA networks as a
    function of $a,$ slope of sigmoid function and $k,$ level of
    desired group ferment. }
  \label{tab:3BA}
\end{table}

\begin{table}[htbp]
  \begin{tabular}{ |P{1.3cm}||P{1.2cm}|P{1.2cm}|P{1.2cm}|P{1.2cm}|  }
    \hline
    $a$ $\downarrow$~~~ $k$ $\rightarrow$ & $0.5$ & $0.6$ & $0.7$ & $ 0.8$\\
    \hline
    $1$  & 253.99&	681.72&	1427.35&	 2762.23 \\
    $3$  & 267.60&	406.50&	594.21&	876.89\\ 
    $5$  & 281.17&	369.05&	477.83&	632.44\\
    $7$  & 289.95&	355.42&	433.38&	542.56 \\
    $10$ & 296.80&	345.79&	403.86&	482.92 \\
    \hline
  \end{tabular}
  \vspace{0.05cm}
  \caption{Cost of maintaining group ferment in $k$R networks as a
    function of $a,$ slope of sigmoid function, and $k,$ level of
    group ferment desired. }
   \label{tab:3KR}
\end{table}

\begin{table}[htbp]
  \begin{tabular}{ |P{1.3cm}||P{1.2cm}|P{1.2cm}|P{1.2cm}|P{1.2cm}|  }
    \hline
    $a$ $\downarrow$~~~ $k$ $\rightarrow$ & $0.5$ & $0.6$ & $0.7$ & $ 0.8$\\
    \hline
    $1$  & 114.06&	301.50&	622.50&	1195.90  \\
    $3$  & 106.94&	160.59&	234.19&	344.85\\ 
    $5$  & 109.76&	144.22&	187.78&	248.73\\
    $7$  & 112.18&	138.62&	170.57&	213.49  \\
    $10$ & 105.56&	124.47&	146.59&	175.17 \\
    \hline
  \end{tabular}
  \vspace{0.05cm}
  \caption{Cost of maintaining group ferment in ER networks as a
    function of parameters $a,$ slope of sigmoid function, and $k,$
    level of group ferment desired. }
  \label{tab:3ER}
\end{table}

In Tables~\ref{tab:3BA},~\ref{tab:3KR}, and~\ref{tab:3ER}, note that
the cost is increasing with increase in the parameter $k$. This is to
be expected because a larger value of $k$ implies a stricter ferment
requirement. For most values of $k$, the cost decreases with an
increase in the slope parameter $a$. A higher value of $a$ implies
that the sigmoid function used in the group ferment requirement
resembles the step function more closely, which in turn means that the
condition can be satisfied by targeting control on to a smaller
fraction of nodes and ignoring the nodes which are more difficult to
influence. It can also be observed that the increments in the cost
with $k$ are larger for smaller values of $a$. This is because a
smaller value of $a$ implies that the sigmoid function has a smaller
slope and a larger value of control is required to achieve the same
increase in $\phi(x(\cdot)).$

We also tabulate the average number of nodes whose opinion values are
actually above the threshold at equilibrium when the control is
applied to maintain group ferment. From
Tables~\ref{tab:4BA},~\ref{tab:4KR}, and~\ref{tab:4ER}, it can be seen
that group ferment can be obtained by taking a smaller number of nodes
above the threshold for smaller values of $k$. However for a higher
value of $k,$ the group ferment is ensured only when a larger number
of nodes are above the threshold, and as seen in
Tables~\ref{tab:3BA},~\ref{tab:3KR}, and~\ref{tab:3ER}, this comes at
an increased cost.  \rev{It can be seen that for some small values of
  $k,$ $a$ needs to be large to achieve the objective of having $kn$
  nodes above $\tau.$ }

\begin{table}[htbp]
  \begin{tabular}{ |P{1.3cm}||P{1.2cm}|P{1.2cm}|P{1.2cm}|P{1.2cm}|  }
    \hline
    $a$ $\downarrow$~~~ $k$ $\rightarrow$ & $0.5$ & $0.6$ & $0.7$ & $ 0.8$\\
    \hline
    $1$  &  17.10 & 50.00 & 50.00 & 50.00  \\
    $3$  &  17.23 & 44.63 & 50.00 & 50.00 \\ 
    $5$  &  17.68 & 39.35 & 47.41 & 50.00 \\
    $7$  &  19.32 & 36.21 & 45.13 & 49.31  \\
    $10$ &  20.06 & 34.01 & 42.24 & 48.03  \\
    \hline
  \end{tabular}
  \vspace{0.05cm}
  \caption{Average number of nodes with opinions above threshold at
    equilibrium in the \eqref{eq:GF-OCP} problem in BA networks, as
    a function of the parameters $a:$ slope of sigmoid function and
    $k:$ the level of group ferment desired. }
   \label{tab:4BA}
\end{table}

\begin{table}[htbp]
\begin{tabular}{ |P{1.3cm}||P{1.2cm}|P{1.2cm}|P{1.2cm}|P{1.2cm}|  }
 \hline
   $a$ $\downarrow$~~~ $k$ $\rightarrow$ & $0.5$ & $0.6$ & $0.7$ & $ 0.8$\\
 \hline
 $1$  &  7.54 & 50.00 & 50.00  & 50.00  \\
 $3$  &  12.70 & 38.61 & 50.00 & 50.00 \\ 
 $5$ & 18.51 & 34.17 &  48.97 &  50.00 \\
 $7$  & 23.11 & 33.54 & 44.46 & 49.91  \\
 $10$  & 27.03 & 33.32 & 37.77 & 48.74  \\
 \hline
\end{tabular}
 \vspace{0.05cm}
\caption{The average number of nodes which are actually above the
  threshold at equilibrium in the \eqref{eq:GF-OCP} problem in $k$R
  networks, as a function of the parameters $a:$ slope of sigmoid
  function and $k:$ the level of group ferment desired. }
   \label{tab:4KR}
\end{table}

\begin{table}[htbp]
\begin{tabular}{ |P{1.3cm}||P{1.2cm}|P{1.2cm}|P{1.2cm}|P{1.2cm}|  }
 \hline
   $a$ $\downarrow$~~~ $k$ $\rightarrow$ & $0.5$ & $0.6$ & $0.7$ & $ 0.8$\\
 \hline
 $1$  &  17.66 & 49.81 & 50.00 & 50.00  \\
 $3$  &  17.58 & 34.91 & 45.58 & 45.83 \\ 
 $5$  &  18.83 & 31.78 & 41.91 & 45.75 \\
 $7$  &  20.49 & 30.31 & 38.33 & 45.35 \\
 $10$ &  19.26 & 26.73 & 32.50 & 39.77  \\
 \hline
\end{tabular}
 \vspace{0.05cm}
\caption{The average number of nodes which are actually above the
  threshold at equilibrium in the \eqref{eq:GF-OCP} problem in ER
  networks, as a function of the parameters $a:$ slope of sigmoid
  function and $k:$ the level of group ferment desired. }
   \label{tab:4ER}
\end{table}

\subsection{Maxmin Ferment ~\eqref{eq:MF-OCP}}
\begin{figure}
  \centering{
    \includegraphics[width=0.8\columnwidth,height
      =0.6\columnwidth]{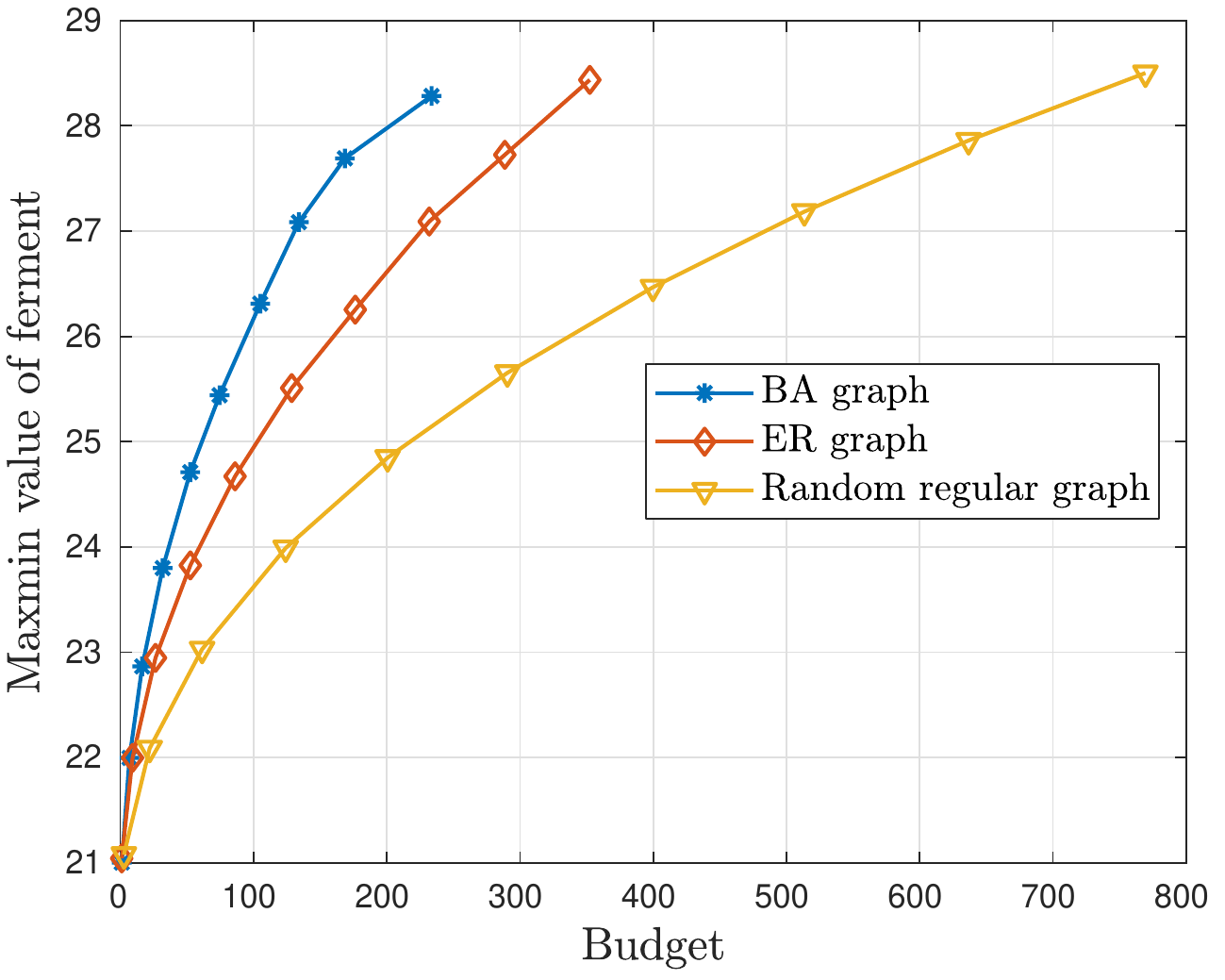} }
  \caption{Ferment level attained as a function of the budget in
    maxmin problem for $k$R, ER, and BA networks.}
  \label{fig:maxmin_all}
\end{figure}

Finally, we study the~\eqref{eq:MF-OCP} problem. Specifically, we plot
the maxmin ferment level attained as a function of the available
budget.  The underlying graph has $n=50$ nodes and the average degree
is $6$. The threshold parameter $\tau$ is set to $0.7$. The system is
observed for $T = 100$ time steps. The initial opinion level is set to
a high value, $\bar{x}_i = 2 ~\forall i \in [1, \ldots, n]$. The slope
parameter $a = 0.5$.
The number of controlled nodes is $m=5$ and they are chosen by the
degree centres heuristic. \rev{For this problem, we choose the degree
  centres heuristic instead of the greedy heuristic because the greedy
  heuristic is based upon the condition that the Turnpike property is
  present in the state-control trajectories, which is not true for
  this problem. On the other hand, the degree centres heuristic is
  more general in definition and scope.} The results corresponding to
BA, ER and $k$R networks are in Fig.~\ref{fig:maxmin_all} where the
value of $\underset{t}{\text{minimum~}} \psi(x(t))$ is plotted as a
function of the budget. Each point on the plot represents the average
ferment and average budget across $50$ observations of the random
graph topologies. \revv{Since we use a iterative algorithm to converge to the intended budget value, we also collect data points corresponding to the sample path and then bin the observations into $10$ groups. The average budget and the average ferment for each bin is plotted.} It can be seen that the ferment is costliest for the
$k$R graph and cheapest for the BA graph. The results corresponding to
larger values of $a$ could not be attained due to stability issues of
the numerical procedure.

\section{Discussion}
\label{sec:discussion}

We considered the control problem of influencing opinion dynamics on a
social network to achieve control objectives. Motivated by `mindshare'\rev{-}like objectives of campaigns, we were interested in the opinion levels
(states) over the entire control period. We considered two types of
problems---a minimum cost problem with constraints on the allowable
states in $\{T_0,\ldots,T\}$ and a maxmin problem with total cost
constraint.

Several open questions remain even within the problem space that we
considered.  Our extensive simulations indicate that the set of
controlled nodes obtained as a solution to the convex relaxation
\eqref{convex_relaxation} for both TF and GF is submodular. If this
were the case then the greedy heuristic is within $(1-1/e)$ of the
optimal value. Also, Algorithm~\ref{algo:maxmin} that determines the
control function for the maxmin problem has numerical stability
issues. An algorithm without this deficiency is desirable. 

There are of course several variations possible both on the opinion
dynamics, and on the state constraints. We expect that the key
techniques and the results would be along similar lines. An important
class of problems would be to consider multiple opposing campaigns
possibly with different control nodes. For example, in \cite{Goyal20},
we have considered two opposing campaigns in a continuous time
$SI_1SI_2S$ epidemic model and obtained the Nash control
strategies. There are of course myriad possibilities. 

\appendix
\section{Appendix - Proofs}

\subsection{Proof of Lemma~\ref{lemma:lem3}}
\label{app:lemma3-proof}
Consider the system
  \begin{align}
    & \text{minimize} ~~ \| u^e \|_2^2 \\ & \text{subject to }
    \begin{cases}
      & x^e = Ax^e + Bu^e + (I_n-A)q   \label{equilibrium}\\
      &  x^e \geq \tau.
    \end{cases}
  \end{align}
  Rewrite~\eqref{equilibrium} as
  \begin{align}
    (I_n -A) (x^e - q) &= Bu^e, \\
    x^e &\geq \tau.
  \end{align}
  \revv{After a change of variable, $(x^e - q) \to x^e$,} we get
  \begin{align}
    (I_n -A) x^e &= Bu^e, \\
    x^e &\geq \tau - q.
  \end{align}
  We will now focus on the matrix $(I_n -A)$ and its inverse. 
  \revv{See that since $A$ is substochastic, $(I_n-A)^{-1} = \sum_{i = 0}^{\infty} A^i$. Since all the entries of $A^i$ are non-negative for all $i \in \{0, \cdots, \infty\}$, $(I_n-A)^{-1}$ has all entries non-negative.} That is:
  \begin{equation}
    (I_n-A)^{-1} \succcurlyeq 0. \label{positive}
  \end{equation} 
  Now consider the following:
  \begin{align}
    x^e &= (I_n-A)^{-1} B u^e, \\
    x^e &\geq \tau - q.
  \end{align}
  Let the set of controlled nodes be $C.$ Here, we have that $Bu^e \in \mathbb{R}^{n \times 1}$ has the $k$-th entry non-zero for all $k \in C.$ Let the non-zero entries of $Bu^e$ be $c_k > 0$ for all $k \in C.$ Denote the $l$-th column of matrix $M$ by $M_l.$ Now, the constraints can be written as:
  \begin{align}
    x^e &= \sum_{k \in C} c_k[(I_n-A)^{-1}]_k \\
    x^e &\geq \tau - q.
  \end{align}
   Since $A$ is sub-stochastic, we have $(I_n-A)^{-1} = \sum_{i = 0}^{\infty} A^i.$ Now the original problem can be re-stated as:
  \begin{align}
    & \text{minimize} ~~~ \sum_{k \in C}  c_k^2 \\
    & \text{subject to } ~ \sum_{k \in C} c_k \left[\sum_{i = 0}^{\infty} A^i \right]_k \geq \tau - q.
  \end{align}
  It follows from \eqref{positive} that the above problem has a
  solution if for all rows of $\sum_{i = 0}^{\infty} A^i$ there is a non-zero entry in at least one of the columns indexed $k$ for $k \in C$. This is ensured by Assumption~\ref{assumption:path_from_controllable}. 
  \subsection{Proof of \eqref{eq:dd}}
  \label{appendix:2}
  Here we give a proof that when at time $t = t_3,$ $r(t_3) =
  \psi(x(t_3)),$ the directional derivatives in direction $v$ are
  equal to~\eqref{eq:dd}.  From~\eqref{eq:ddH}, we have
  \begin{align}
    &\mathcal{D}_v H(\lambda,\cdot,\cdot,y,u)(xr) \nonumber\\    
    &= \lim_{\theta \downarrow 0}\frac{ H(x+\theta v_x, r+\theta v_r,
      y, u) - H(x, r, y, u)}{\theta} \nonumber \\    
    &= \lim_{\theta \downarrow 0}\frac{\lambda_x^{\intercal}(t_3)
      A(x(t_3)+ \theta v_x -x(t_3))}{\theta} \nonumber \\    
    &+ \lim_{\theta \downarrow 0}\frac{\lambda_r(t_3)(
      \text{min}\{\psi(x(t_3)+\theta v_x), r(t_3)+\theta v_r\} -
      r(t_3))}{\theta} \nonumber \\   
    &=\lambda_x^\intercal(t_3) A v_x + \lambda_r(t_3) \min\left\{
    \left< \frac{d}{dx}\psi(x(t_3)),v_x \right>, v_r
    \right\}. \nonumber
  \end{align}

\bibliographystyle{plain}        
\bibliography{cit}
\end{document}